\documentclass[a4paper,11pt,twoside]{article}
\usepackage{amsmath,amssymb,amsthm}
\usepackage[ps,dvips,frame]{xy}

\newcommand{\im}{\mathrm{i}}  
\newcommand{\defeq}{:=}

\newcommand{\N}{\mathbb{N}}
\newcommand{\Z}{\mathbb{Z}}

\newcommand{\R}{\mathbb{R}}
\newcommand{\C}{\mathbb{C}}
\newcommand{\tens}{\otimes}  

\newcommand{\xd}{\mathrm{d}}  

\newcommand{\lalg}[1]{\mathfrak{#1}}  
\newcommand{\ealg}{\mathcal{U}}  
\newcommand{\falg}{\mathcal{C}}  
\DeclareMathOperator{\id}{id}

\newcommand{\tos}{\twoheadrightarrow}
\newcommand{\toi}{\hookrightarrow}

\newcommand{\one}{\mathbf{1}}

\DeclareMathOperator{\cou}{\epsilon}
\DeclareMathOperator{\cop}{\Delta}

\DeclareMathOperator{\antip}{\mathrm{S}}

\newcommand{\act}{\triangleright}

\renewcommand{\i}[1]{{}_{\scriptscriptstyle(#1)}}
\newcommand{\iu}[1]{{}_{\scriptscriptstyle(\underline #1)}}
\newcommand{\ib}[1]{{}_{\scriptscriptstyle[#1]}}
\newcommand{\ibu}[1]{{}_{\scriptscriptstyle[\underline #1]}}

\newcommand{\SU}[1]{SU(#1)}

\newcommand{\catmodcl}[1]{{}^{#1}{\mathcal{M}}}

\newcommand{\mR}{\mathsf{R}}
\newcommand{\cR}{\mathcal{R}}

\theoremstyle{plain}
\newtheorem{prop}{Proposition}[section]
\newtheorem{lem}[prop]{Lemma}
\newtheorem{thm}[prop]{Theorem}
\newtheorem{cor}[prop]{Corollary}
\newtheorem{ex}[prop]{Example}
\newtheorem{dfn}[prop]{Definition}
\newtheorem{rem}[prop]{Remark}

\newcommand{\Pc}{\hat{P}}
\newcommand{\qg}[1]{\mathsf{#1}}

\newcommand{\rbos}{\rtimes}

\newcommand{\CZ}[1]{\qg{Z_{#1}}'}     
\newcommand{\CtZ}[1]{\qg{Z_{#1}}}     
\newcommand{\CSU}[1]{\qg{SU}(#1)}
\newcommand{\CgSU}[1]{\qg{SU}'(#1)}
\newcommand{\Bt}{\tilde{B}}
\newcommand{\sigmat}{\tilde{\sigma}}

\newcommand{\mom}{\lalg{tr_4}}
\newcommand{\smom}{\lalg{str_4}}
\newcommand{\llor}{\lalg{so_{3,1}}}
\newcommand{\mink}{\mathsf{Mink}}
\newcommand{\smink}{\mathsf{SMink}}
\newcommand{\lpoinc}{\lalg{poinc}}
\newcommand{\lspoinc}{\lalg{spoinc}}
\newcommand{\poinc}{\mathsf{Poinc}}
\newcommand{\poincs}{\mathsf{Poinc'}}
\newcommand{\spoincs}{\mathsf{SPoinc'}}
\DeclareMathOperator{\tr}{tr}

\newcommand{\sxy}[1]{{\begin{xy} #1 \end{xy}}}
\everyxy={0;<1mm,0mm>:<0mm,1mm>::0;0,}

\allowdisplaybreaks

\begin{document}
\title{\textbf{The Quantum Geometry of Supersymmetry and the
Generalized Group Extension Problem}}
\author{Robert Oeckl\footnote{email: oeckl@cpt.univ-mrs.fr}\\ \\
Centre de Physique Th\'eorique,
CNRS Luminy, Case 907,\\
F-13288 Marseille -- Cedex 9}
\date{CPT-2001/P.4212\\
14 June 2001, 22 March 2002 (v2)}

\maketitle

\vspace{\stretch{1}}

\begin{abstract}
We examine the notion of symmetry in quantum field
theory from a fundamental representation theoretic point of view.
This leads us to a generalization expressed
in terms of quantum groups and braided categories.
It also unifies the conventional concept of symmetry with that
of exchange statistics and the spin-statistics relation.
We show how this quantum group symmetry is reconstructed from the
traditional (super) group symmetry, statistics and spin-statistics
relation.

The old question of extending the Poincar\'e group to unify external
and internal symmetries (solved by supersymmetry) is reexamined in the
new framework.
The reason why we should allow supergroups in this case
becomes completely transparent.
However, the true symmetries are not expressed by groups or
supergroups here but by ordinary (not super) quantum groups.
We show in this generalized framework that
supersymmetry remains the most general unification of internal and space-time
symmetries provided that all particles are either bosons or fermions.

Finally, we demonstrate with some examples how quantum geometry
provides a natural setting for the construction of super-extensions,
super-spaces, super-derivatives etc.

\end{abstract}

\vspace{\stretch{1}}

\clearpage
\section{Introduction}

The question was raised a long time ago whether the external
(space-time) and internal symmetries of the quantum field theories
with which we describe nature
could be part of a larger symmetry group that is not simply a direct
product of the two.

For relativistic quantum mechanical theories the space-time
symmetry group is the universal cover $\Pc$ of the Poincar\'e group
$P$. (For simplicity we refer to $\Pc$ in the following as the
Poincar\'e group.)
Thus, a unification of symmetries in the abovementioned sense
would imply a solution to the following problem:
Is there a larger group $S\Pc$ which contains the
Poincar\'e group
$\Pc$, but is not simply a direct product of $\Pc$ and some other
group? That is, is there a group $S\Pc$ with an inclusion
\begin{equation}
 \Pc\hookrightarrow S\Pc\quad\text{such that}\quad S\Pc\neq \Pc\times G
\label{eq:gxprob}
\end{equation}
for any group $G$?

While mathematical solutions to the problem in this simple form can be
easily found,
they might not be of physical relevance.
One can enlarge $\Pc$ for example by adding scale transformations.
However, scale invariance is not a feature of the physically
relevant quantum field theories of fundamental interactions.
One therefore needs to impose additional constraints on
$S\Pc$ in order for it
to be physically interesting.
Precisely such an analysis was carried out in the context of
scattering theory in the 1960's, and brought
into its most comprehensive form by
Coleman and Mandula \cite{CoMa:allsymsmat}.
They were able to show that under 
reasonable physical assumptions the Lie algebra version of problem
(\ref{eq:gxprob}) has no solution: There is no such extension
of the Poincar\'e Lie algebra.

Only a few years later, however,
supersymmetry emerged as a physically acceptable solution
to the extension problem in a modified from
\cite{GoLi:extalgpoinc,WeZu:suga}.
One needs to extend the concept of symmetry from that of groups and
Lie algebras to that of supergroups and super-Lie algebras.
Then, a  physically acceptable extension of the Poincar\'e Lie algebra
exists: The super-Poincar\'e Lie algebra.
The analysis of Coleman and Mandula was repeated
by Haag, \L{}opusza\'nski, and Sohnius for the super-Lie algebra
case \cite{HaLoSo:allsusysmat}. They found the super-Poincar\'e Lie
algebra (in
its versions with various numbers of supersymmetries and additional
central charges) to be the only 
physically acceptable extension of the Poincar\'e Lie
algebra.

Is this the end of the story?
Can we go beyond supergroups and supersymmetry? And why ``super'' in
the first place?

In the following we try to answer these questions from a
categorial (or representation theoretic) point of view.
This leads us to a unified view of symmetry and statistics through
braided categories and
quantum groups (in Section~\ref{sec:qgsym}). This generalized notion
of symmetry then provides the natural framework for posing the
analogue of the
extension problem (\ref{eq:gxprob}) (in Section~\ref{sec:qextprob}).
In Section~\ref{sec:essqg} we introduce the reader to the necessary
essentials from quantum group theory and provide some elementary
examples.
Section~\ref{sec:recpoinc} is devoted to reconstructing the quantum
group symmetry underlying ordinary quantum field theory. As it turns
out this is not the ordinary Poincar\'e group but a closely related
quantum group. The reconstruction is then generalized (in
Section~\ref{sec:frec}) and applied to the extension problem (in
Section~\ref{sec:recext}). The latter section provides the link
between the superextension problem and our generalized extension
problem. In Section~\ref{sec:nobeyond} we pursue the question of
whether (for ordinary QFT) there is something
``beyond supersymmetry''. The answer is ``No'' and indeed the main
mathematical result here is that all possible extensions (in the case
of Bose-Fermi statistics) can be obtained from groups or supergroups.

As it turns out, our setting also provides us
with new mathematical tools for dealing with supersymmetry.
These are the tools of quantum geometry \cite{Maj:qgroups}. By quantum
geometry we mean
here the noncommutative
geometry whose manifold-objects are algebras and whose
group-objects are quantum groups (Hopf algebras).
In quantum geometry there are generalizations of principal bundles,
homogeneous spaces, differential forms etc.
We give examples in Section~\ref{sec:exappl} of how all this can be
applied to supersymmetry and facilitates supersymmetric
constructions.
These include semidirect superextensions, the OSp-supergroups,
super-spheres and the super-Poincar\'e group.

Proofs for mathematical statements are in general omitted as they are
either known or straightforward. In the former case either a reference
is given or they can be found in text books on quantum groups.
An exception forms Theorem~\ref{thm:qexttogext} whose proof is
explicitly given.

We work throughout over the field of complex numbers.
\section{The Generalized Extension Problem}

\subsection{Why Quantum Group Symmetries?}
\label{sec:qgsym}

With the insufficiency of the group context in mind, we search for
a more general but natural framework for the notion of symmetry
and the extension problem.
We are hereby guided by the categorial (i.e.\ representation theoretic)
aspects of quantum field theory.

What are the ``objects'' that we deal with?
States, fields, operators, Lagrangians etc.\ all live in
vector spaces over $\R$ or $\C$. Furthermore, they all carry
actions of the Poincar\'e group $\Pc$ or some larger symmetry
group of the theory. That is, these vector spaces are representations
of the symmetry group. Furthermore, there are maps between the
representations which are required to be intertwiners, i.e., they
commute with the group action. For example, an invariant operator can
be viewed as such a map between states. 
What we have described so far, objects and maps between them,
is essentially what makes a \emph{category}. In this case, it is the
category of representations of the symmetry group.

An essential operation in quantum field theory is the formation
of tensor products of representations, e.g., to form
a two-particle state out of two one particle states.
This gives additional structure to the category of representations
of the symmetry group and makes it into a \emph{monoidal category}.
In fact, this monoidal category carries all the information about
the representation theory and we can forget about
the group itself altogether.

We already know that we need to generalize the symmetry concept beyond
that of groups to allow for supersymmetry. However, replacing
groups by supergroups leads to monoidal categories as
well. Conversely, given a monoidal category we require
no knowledge about an underlying group or supergroup to perform
all the representation theoretic operations necessary in quantum
field theory.
Thus, it appears natural to define a generalized concept of symmetry
simply by that of a monoidal category.

However, there is a theorem of quantum group theory that states
that for any
monoidal category (with duals) there is a Hopf algebra so that the
monoidal category is its category of representations.\footnote{We use
the word ``representation'' for a Hopf algebra here and in the
following to mean ``comodule''. See Section~\ref{sec:essqg} for more
details.}
This is called Tannaka-Krein reconstruction (see \cite{Maj:qgroups}).
In fact, this gives rise to a one-to-one correspondence between monoidal
categories and Hopf algebras. Thus, the abstract generalization to
any monoidal category gives us back a more concrete object 
that encodes the symmetries --
a Hopf algebra. In the group case, this Hopf algebra is the
commutative Hopf algebra of functions on the group.\footnote{We are
somewhat sloppy here and in the following concerning functional
analytic questions such as the choice of class of functions on a space
or the necessity to complete tensor products, consider multiplier
algebras etc. The treatment of these questions would unnecessarily
complicate the discussion and is irrelevant for the purposes of this
paper.}
In the supergroup case
the relation to the corresponding Hopf algebra is
slightly more complicated (see Section~\ref{sec:rec}).

We can go on to exploit our categorial point of view further
to encompass the notion of particle statistics as well.
In fact, this turns out to be \emph{essential} as symmetry and
statistics
become inseparably linked in the generalized Hopf algebraic context.

A bit less obvious, it is also an essential ingredient of quantum field
theory to have for two representations
$V$ and $W$ an intertwiner $V\tens W\to W\tens V$. For two one-particle
states this intertwiner tells us what the exchange statistics of
the particles is. E.g., for Bosons this would be
$v\tens w\mapsto w\tens v$ while for Fermions we would have an extra
minus sign $v\tens w\mapsto -w\tens v$. In general, the definition
of such an intertwiner for any pair of representations is called
a \emph{braiding}. Thus, the objects of a quantum field
theory live in a \emph{braided monoidal category}.
This encodes now both, the symmetries and the statistics
of the theory. Note that this concept allows for more general
statistics than Bose and Fermi, see \cite{Oe:spinstat} for a
discussion.

The braiding on the category as a category of representations yields
an extra structure on the corresponding Hopf algebra
via Tannaka-Krein reconstruction. 
This is called a \emph{coquasitriangular structure}.
Again, this gives rise to a one-to-one
correspondence between braided monoidal categories and
coquasitriangular Hopf algebras.
In the following, we use the term \emph{quantum group} to denote
coquasitriangular Hopf algebras.

Importantly, it is not possible to combine arbitrary Hopf algebras
with arbitrary braidings. To the contrary, for a given Hopf algebra
the set of possible braidings on its representation category
(encoded in the coquasitriangular structure) is usually very limited.
Thus, symmetry and statistics cannot be viewed as separate entities
in general. We subsume both under a generalized notion of symmetry
which replaces ordinary groups by quantum groups.
Unsurprisingly, also supersymmetry gives rise to a particular
example of such a generalized symmetry, as we shall discuss
in Section~\ref{sec:recext}.

\subsection{The Quantum Group Extension Problem}
\label{sec:qextprob}

Let us examine the extension problem (\ref{eq:gxprob}) from the same
abstract representation theoretic point of
view that we have employed in the previous section.

Suppose we wish to embed a group $G$ into a larger group $G'$.
That is, we look for an inclusion $G\hookrightarrow G'$.
For the moment suppose we are just given a group
homomorphism $G\to G'$.
For the representations this means that we
can pull back a representation of $G'$ to one of $G$.
In fact, this gives rise to
a (monoidal) functor between the (monoidal) categories of
representations of the groups in the opposite direction
${}_{G'}\mathcal{M}\to{}_G\mathcal{M}$. That is, for every
representation of $G'$ we get one of $G$ and for every intertwiner
between representations of $G'$ we get one between representations of
$G$.
Conversely, given this functor
we can reconstruct the group homomorphism. Indeed, there is a
one-to-one
correspondence between such functors and group homomorphisms.

Generalizing as in the previous section
to the case of arbitrary monoidal categories we still
have such a correspondence. It is between functors and Hopf algebra
homomorphisms. This time, both arrows point in the same
direction. Thus, the generalization of the group
homomorphism $G\to G'$ is a Hopf algebra homomorphism $H'\to H$.
We recover the group case from the Hopf algebra case with
the function Hopf algebras $H=\falg(G)$, $H'=\falg(G')$.
The injectivity of the group homomorphism corresponds to the
surjectivity of the Hopf algebra homomorphism. Thus, the problem
of finding a ``larger'' group $G'$ in which to embed a group $G$
generalizes to the problem of finding a ``larger'' Hopf algebra $H'$
with a surjection $H'\twoheadrightarrow H$ to the given Hopf algebra $H$.

While in the group extension problem (\ref{eq:gxprob}) the exchange
statistics is not explicitly mentioned and only enters separately
in the physical conditions we can do better with our generalized
setting of Section~\ref{sec:qgsym}. To include the statistics we only
have to consider the braiding that encodes it as well.
Thus, we have braided monoidal categories instead of just
monoidal categories.
For a (monoidal) functor between such categories
we impose the natural condition of being braided, i.e., of
commuting with the braiding.
This exactly expresses the condition that the statistics
is preserved by the extension.
We then have
a correspondence between braided monoidal functors
and homomorphisms of coquasitriangular Hopf algebras
(quantum groups). Thus, the extension problem becomes that of finding
a ``larger'' quantum group $H'$ with a surjection (of coquasitriangular
Hopf algebras) $H'\twoheadrightarrow H$ to a given quantum group.

The analogue of the condition that the ``larger'' group
not be a direct product corresponds to the ``larger'' quantum
group not being a tensor product. Thus, we can formulate the
quantum group
generalization of (\ref{eq:gxprob}) as follows:
Denoting the relevant quantum group version of the Poincar\'e group by
$\qg{\Pc}'$,\footnote{$\qg{\Pc}'$ encodes now the Poincar\'e
symmetry as well as Bose-Fermi statistics and the spin-statistics
relation. It is derived in Section~\ref{sec:recpoinc}.}
find a quantum group $\qg{S\Pc}'$ and a surjection
\begin{equation}
 \qg{S\Pc}'\twoheadrightarrow \qg{\Pc}'\quad\text{such that}\quad
 \qg{S\Pc}'\neq \qg{\Pc}'\tens\qg{G}
\label{eq:qxprob}
\end{equation}
for any quantum group $\qg{G}$.

\section{Essentials from Quantum Group Theory}
\label{sec:essqg}

In this section we introduce a few essential elements of quantum group
theory and give some elementary examples. The latter serve to
acquaint the reader with the formalism and form at the same time the
basis for supersymmetric examples in Section~\ref{sec:exappl}.
Most of the material in this section is text book knowledge. A good
reference is Majid's book \cite{Maj:qgroups}, in particular for the
braided aspects. For the material on specific groups and Lie
algebras see e.g.\ \cite{Cor:groupphys12}.

We assume the reader to be familiar with the notions of Hopf algebra,
module, comodule, and Hopf algebra pairing.
We use the notations $\cop,\cou,\antip$ for coproduct,
counit and antipode of a Hopf algebra. We use
Sweedler's notation (with implicit summation) $\cop a=a\i1\tens a\i2$
for coproducts and a
similar notation $v\mapsto v\i1\tens v\iu2$ for left coactions.

A braided monoidal category is a monoidal category (i.e.\ a collection of
objects and maps with a tensor product and certain compatibility
conditions)
so that for any two
objects $V,W$ there is an invertible map $\psi:V\tens W\to W\tens V$
(the \emph{braiding}).
The collection of $\psi$'s also has to satisfy certain compatibility
conditions. A braiding is called \emph{symmetric} if $\psi=\psi^{-1}$.

A \emph{coquasitriangular} structure $\cR:H\tens H\to\C$ on a Hopf algebra
$H$ provides a braiding on its category of left comodules via
\[
 \psi(v\tens w)=\cR(w\i1\tens v\i1)\, w\iu2\tens v\iu2 .
\]
If $\cR(a\i1\tens b\i1) \cR(b\i2\tens a\i2)=\cou(a)\cou(b)$, then
$\cR$ is called \emph{cotriangular} and the induced braiding is
symmetric.

Dually, a \emph{quasitriangular} structure $\mR\in H\tens H$ on a Hopf
algebra $H$ provides a braiding on its category of left modules via
\[
 \psi(v\tens w)=\mR_{2}\act w \tens \mR_{1}\act v
\]
with $\mR_1\tens \mR_2\defeq\mR$ (summation implied). If
$\mR^{-1}=\mR_2\tens \mR_1$, then $\mR$ is called \emph{triangular}
and the induced braiding is symmetric.

As alluded to above, a group $G$ gives rise to a Hopf algebra
as follows. Take the algebra of functions $\falg(G)$ on $G$ and equip
it with a coproduct defined by $(\cop f)(g,h)=f(g h)$ for
$f\in\falg(G)$ and $g,h\in G$
using the
identification $\falg(G\times G)\cong \falg(G)\tens\falg(G)$. Counit
and antipode are given by $\cou(f)=f(e)$ and $(\antip f)(g)=f(g^{-1})$
where $e$ denotes the unit element of the group. Note that the Hopf
algebra $\falg(G)$ naturally carries the trivial cotriangular
structure $\cR=\cou\tens\cou$ which encodes the trivial braiding
$v\tens w\mapsto w\tens v$.

For matrix groups the corresponding Hopf algebra can be constructed
rather explicitly.
Consider the coalgebra with basis $\{t_{ij}\}$ for $i,j\in
\{1,\dots,n\}$, with coproduct $\cop t_{ij}=\sum_k t_{ik} \tens
t_{kj}$ and counit
$\cou(t_{ij})=\delta_{ij}$. It is called the $n$-dimensional \emph{matrix
coalgebra} and is dual to the algebra of $n\times n$-matrices $M_n$. The
free commutative bialgebra generated by the $t_{ij}$ is the
``prototype'' of the function Hopf algebra of a matrix group. More
precisely, a matrix group that is a subalgebra of $M_n$ determined by
polynomial constraints corresponds to a Hopf algebra which is a
quotient of the described bialgebra by relations corresponding to the
constraints.

From here on we adopt the convention that we denote the Hopf algebra
of functions on a group $G$ by $\qg{G}$. The class of functions we
usually choose are the representative functions. These are the
functions that arise as matrix elements of finite-dimensional
representations. Furthermore we sometimes consider a
\emph{conjugation} in this context. This is nothing but ordinary
complex conjugation.

\begin{ex}
\label{ex:csu2}
Consider the group $SU(2)$. Its Hopf algebra $\qg{SU}(2)$ of
representative functions
is generated by the matrix coalgebra $\{t_{ij}\}$
with $i,j\in\{\frac{1}{2},-\frac{1}{2}\}$
and relation (with the notation $\pm$ for $\pm\frac{1}{2}$)
\[
 t_{--} t_{++}-t_{-+} t_{+-}= 1.
\]
$\qg{SU}(2)$ has a (Peter-Weyl) basis $\{t^l_{i j}\}$ with $l\in
\frac{1}{2}\N_0$ 
and $i,j\in\{-l,-l+1,\dots l\}$ with $t^0_{0 0}=1$ and
$t^{1/2}_{i j}=t_{i j}$.
Coproduct, counit, antipode and conjugation in this basis are given
by 
\[
 \cop t^l_{m n}=\sum_k t^l_{m k} \tens t^l_{k n},\quad
 \cou (t^l_{m n})=\delta_{m n},\quad
  \overline{t^l_{m n}}=\antip t^l_{n m}= (-1)^{n-m} t^l_{-m,-n} .
\]
\end{ex}

\begin{ex}
\label{ex:clor}
Consider the group $Spin(3,1)=SL(2,\C)$ which is the double cover of the
Lorentz group $SO(3,1)$. Its Hopf algebra $\qg{Spin}(3,1)$ of
representative functions is the tensor product of two copies of
$\qg{SU}(2)$ whose generators we denote by $\{t_{i j}\}$ and
$\{\overline{t_{i j}}\}$. However its conjugation is different, as
indicated by the notation for the generators.
A Peter-Weyl basis\footnote{The term ``Peter-Weyl basis''
refers here (as above) to a decomposition in terms of irreducible
finite-dimensional representations and does \emph{not} involve
unitarity in any way.} is thus given
by $\{t^l_{i j},\overline{t^l_{i j}}\}$. 

Let $\{\sigma^\mu\}$ denote the standard Pauli matrices
\[
 \sigma^0\defeq\begin{pmatrix} 1 & 0 \\ 0 & 1 \end{pmatrix},\quad
 \sigma^1\defeq\begin{pmatrix} 0 & 1 \\ 1 & 0 \end{pmatrix},\quad
 \sigma^2\defeq\begin{pmatrix} 0 & -\im \\ \im & 0 \end{pmatrix},\quad
 \sigma^3\defeq\begin{pmatrix} 1 & 0 \\ 0 & -1 \end{pmatrix} .
\]
Define the $2\times 2$ matrix $T$ of generators
by
\[
T\defeq \begin{pmatrix} t_{- -} & t_{- +}\\ t_{+ -} & t_{+ +}
\end{pmatrix}
\]
and the elements
\[
 \Lambda^{\mu \nu}\defeq\frac{1}{2}\tr(\sigma^\mu T \sigma^\nu
 T^\dagger) ,
\]
where
$T^\dagger$ is transposition of the matrix composed with
conjugation of its elements. $\{\Lambda^{\mu \nu}\}$ generates precisely
the sub-Hopf algebra $\qg{SO}(3,1)$ of functions on the Lorentz group.
Note
that $\overline{\Lambda^{\mu \nu}}=\Lambda^{\mu \nu}$.

The surjection $\qg{Spin}(3,1)\tos\qg{SU}(2)$ corresponding to the
injection $SU(2)\toi Spin(3,1)$ is simply given by $t^l_{i j}\mapsto
t^l_{i j}$ and $\overline{t^l_{i j}}\mapsto \overline{t^l_{i
j}}$. (Note the different meaning of the conjugation in source and
target.)
\end{ex}

Not only a Lie group, but also a Lie algebra $\lalg{g}$ gives rise to
a Hopf algebra. More precisely, its universal envelope
$\ealg(\lalg{g})$ can be made into a Hopf algebra. This is achieved by
equipping the Lie algebra generators with the \emph{primitive}
coproduct $\cop\eta=\eta\tens 1+1 \tens\eta$. This determines a
coproduct on the whole of $\ealg(\lalg{g})$. The counit is given by
$\cou(\eta)=0$ on the generators and the antipode by
$\antip\eta=-\eta$. Note that $\ealg(\lalg{g})$ is cocommutative.
Furthermore, we sometimes consider a
conjugation. Then the Lie algebra can be considered as the
complexification of a real Lie algebra with the given complex
conjugation.

It is a remarkable fact of Hopf algebra theory that the Hopf algebras
obtained from a Lie group and its Lie algebra are dual to each other.

\begin{ex}
\label{ex:usl2}
Consider the Lie algebra $\lalg{su_2}$ with basis $E,F,H$ and
relations $[H,E]=2E$, $[H,F]=-2F$, $[E,F]=H$. Conjugation is
given by $\overline{H}=-H$, $\overline{E}=-F$, $\overline{F}=-E$.
Its universal enveloping
Hopf algebra $\ealg(\lalg{su_2})$ is dually paired with $\qg{SU}(2)$ via
\begin{gather*}
 \langle H, t^l_{m n}\rangle=2n\delta_{m,n},\quad
 \langle E, t^l_{m n}\rangle=\sqrt{(l-n)(l+n+1)}\delta_{m,n+1},\\
 \langle F, t^l_{m n}\rangle=\sqrt{(l+n)(l-n+1)}\delta_{m,n-1} .
\end{gather*}
\end{ex}

\begin{ex}
\label{ex:so31}
Consider the Lie algebra $\lalg{so_{3,1}}$ with basis
$E,F,H,\overline{E},\overline{F},\overline{H}$. Apart from the
different conjugation (indicated in the basis) it has the relations of
$\lalg{su_2}\oplus \lalg{su_2}$ in the obvious way. The pairing of
$\ealg(\lalg{so_{3,1}})$ with
$\qg{Spin}(3,1)$ is as in Example~\ref{ex:usl2} for the un-barred and
the same for the barred generators. The pairing between un-barred and
barred generators is zero.

For the elements $\Lambda^{\mu \nu}$ the pairing comes out as
\begin{gather*}
 \langle X, \Lambda^{\mu \nu}\rangle = \frac{1}{2} \tr(\sigma^\mu
 \sigma(X)\sigma^\nu),\quad \langle \overline{X},
 \Lambda^{\mu \nu}\rangle = \overline{\langle X, \Lambda^{\mu
 \nu}\rangle} \quad\forall X\in \{H,E,F\}, \\
 \text{where}\quad \sigma(H)\defeq-\sigma^3,\quad
 \sigma(E)\defeq\frac{1}{2}(\sigma^1-\im\sigma^2), \quad
 \sigma(F)\defeq\frac{1}{2}(\sigma^1+\im\sigma^2) .
\end{gather*}

The injection $\lalg{su_2}\toi \lalg{so_{3,1}}$ corresponding by
duality to the surjection of Example~\ref{ex:clor} is given by
$E\mapsto E-\overline{F}$, $F\mapsto F-\overline{E}$,
$H\mapsto H-\overline{H}$. It extends to $\ealg(\lalg{su_2})\toi
\ealg(\lalg{so_{3,1}})$.
\end{ex}

The simplest example of a Hopf algebra with non-trivial cotriangular
structure (i.e.\ implying non-trivial braiding) is
the following one.

\begin{ex}
\label{ex:z2}
Let $\CZ2$ be the Hopf algebra of functions on $\Z_2$. It has
two elements $1,g$ with relation $g^2=1$, coproduct $\cop g=g\tens g$,
counit $\cou(g)=1$, and antipode $\antip g=g$. We equip it with the
cotriangular structure determined by $\cR(g\tens g)=-1$.
\end{ex}

$\CZ2$ is precisely the quantum group that generates the category of
$\Z_2$-graded vector spaces as its category of comodules. A comodule
$V$ of $\CZ2$ splits into a direct sum $V_0\oplus V_1$ of even and
odd part determined by the coaction $v\mapsto g^{|v|}\tens v$.
This is the natural setting for supergroups and super-Lie
algebras. We start with more general definitions.

A \emph{braided Hopf algebra} is the analogue of a Hopf algebra in a
braided category. That is, it obeys the same axioms as an ordinary
Hopf algebra except for the
axiom of compatibility between product and coproduct which is modified
to
\[
 \cop\circ\cdot = (\cdot\tens\cdot)\circ
 (\id\tens\psi\tens\id)\circ(\cop\tens\cop) .
\]

\begin{dfn}
\label{def:bcom}
Let $A$ be an algebra in a braided category. $A$ is called
\emph{braided commutative} if $\cdot=\cdot\circ\psi$ is an
identity of maps $A\tens A\to A$.

Dually, let $C$ be a coalgebra in a braided category. $C$ is called
\emph{braided cocommutative} if $\cop=\psi\circ\cop$ is an identity of
maps $C\to C\tens C$.
\end{dfn}

Now, in the same way as ordinary Hopf algebras describe groups and
enveloping algebras, $\Z_2$-graded Hopf algebras (i.e., braided Hopf
algebras in the category of $\CZ2$-comodules) describe supergroups
and super-Lie algebras. Thus supergroups are described by
$\Z_2$-graded commutative Hopf algebras and super-enveloping algebras
by $\Z_2$-graded cocommutative Hopf algebras. The latter case is more
familiar. One usually considers super-Lie algebras. Those give indeed
rise to enveloping Hopf algebras precisely in the same way as ordinary
Lie algebras do, except that everything takes place in the $\Z_2$-graded
category.
Our definition of supergroups might seem less familiar but
is standard in the quantum groups literature (see e.g.\
\cite{Man:qdefsuper} where
even $q$-deformations of supergroups were considered). It is also
much less complicated than analytically inspired definitions using
auxiliary Grassmann algebras. 

In fact, one can generalize these considerations to arbitrary braiding
employing the notion of braided Hopf algebra mentioned above, see
\cite{Maj:qgroups}. However, we shall limit ourselves mostly to the
$\Z_2$-graded case, occasionally generalizing to arbitrary symmetric
braidings. For non-symmetric braidings additional problems occur, most
notably the absence of an analogue of Proposition~\ref{prop:ctrbos}
(as discussed in Section~\ref{sec:frec}).

We are now ready to define the extension problems more precisely
(disregarding for now the requirement that the extension must not be a
direct product respectively tensor product).
The conventional version (\ref{eq:gxprob}), generalized to encompass
e.g.\ supergroups in the abovementioned sense can be formulated as
follows:

\begin{dfn}
\label{def:gext}
Let $H$ be a braided commutative Hopf algebra in a symmetric
braided category. Then, the \emph{triangular group extension
problem} is the problem to find a braided commutative Hopf algebra
$B$ in the category with a surjection $\sigma:B\tos H$.
Any such $B$ is said to be a \emph{solution} of the problem.
\end{dfn}

In the ordinary group case the underlying category is just the
category of vector spaces and the braiding is simply the
interchange of the tensor components. $H$ is thus an ordinary
commutative Hopf algebra which encodes the algebra of functions on
a group. In the supergroup case
the category is that of $\Z_2$-graded vector spaces and
the braiding is the interchange with an additional minus sign
if both components are odd. Thus, $H$ is a graded commutative
Hopf algebra which encodes the algebra of functions on a
supergroup.

The quantum group extension problem (\ref{eq:qxprob}) takes the
following form:

\begin{dfn}
\label{def:qext}
Let $H$ be a coquasitriangular Hopf algebra. Then, the \emph{quantum
group extension problem} is the problem to find a coquasitriangular
Hopf algebra $A$ with a surjection $\pi:A\tos H$.
Any such $A$ is said to be a \emph{solution} of the problem.
\end{dfn}

For the following discussion of reconstruction we require the analogue
of a semidirect product of groups for Hopf algebras. This is provided
by the following theorem and its variants. Their significance will
become clear in the next section.

\begin{thm}[Majid \cite{Maj:cpbos,Maj:bmatrix}]
\label{thm:cqtrbos}
Let $H$ be a coquasitriangular Hopf algebra, $B$ a braided Hopf
algebra in the braided category $\catmodcl{H}$ of left
$H$-comodules. Then, there exists a Hopf algebra $B\rbos H$, called
the \emph{bosonization} of $B$, such that the category of left
comodules of $B\rbos H$ is monoidal equivalent to the category of
braided left comodules of $B$ in $\catmodcl{H}$.

Explicitly, $B\rbos H$ is isomorphic to $B\tens H$ as a vector
space. Its product, coproduct, and antipode are given by
\begin{gather*}
 (b\tens h) (c\tens g) = \cR(c\ib1\tens h\i1)\, b c\ibu2\tens h\i2 g ,\\
 \cop (b\tens h) = (b\i1 \tens b\i2\ib1 h\i1)
  \tens (b\i2\ibu2 \tens h\i2) ,\\
 \antip (b\tens h)= \cR((\antip_B b\ibu2)\ib1\tens
  (\antip_H (b\ib1 h))\i1)\,
  (\antip_B b\ibu2)\ibu2
  \tens (\antip_H (b\ib1 h))\i2
\end{gather*}
for $b,c\in B$ and $h,g\in H$.
Here, the coaction of $H$ on $B$ is denoted by $b\mapsto b\ib1\tens
b\ibu2$.

Furthermore, there is a Hopf algebra surjection $\pi:B\rbos H\tos H$
defined by $b\tens h\mapsto\cou(b) h$ and an injection $i:H\toi B\rbos
H$ defined by $h\mapsto 1\tens h$ such that $\pi\circ i=\id$.
Conversely, let $\pi:A\tos H$ be a Hopf algebra surjection. Then $A$
is a bosonization $A=B\rbos H$ for some $B$ if and only if there is an
injection $i:H\toi A$ such that $\pi\circ i=\id$.
\end{thm}

\begin{prop}
\label{prop:ctrbos}
In the context of Theorem~\ref{thm:cqtrbos} assume that $H$ is
cotriangular and that $B$ is braided commutative. Then, $B\rbos H$
inherits a cotriangular structure from $H$ by pull-back, i.e.,
\[
 \cR((b\tens h) \tens (c\tens g))\defeq \cR_H(h\tens
g)\cou(b)\cou(c) .
\]
Furthermore the equivalence of categories of
Theorem~\ref{thm:cqtrbos} becomes an equivalence of braided categories
in this way. In particular, $B\rbos H\tos H$ is a quantum group
extension in the sense of Definition~\ref{def:qext}.
\end{prop}

For the pairing of bosonizations we need the following lemma.

\begin{lem}
\label{lem:pbos}
Let $A$ be a coquasitriangular Hopf algebra and 
$H$ a quasitriangular Hopf algebra which are dually
paired via $H\tens A\to\C$. Let $B$ be an $A$-comodule braided Hopf
algebra and $D$ an $H$-module braided Hopf algebra such that they are
dually paired as algebra/coalgebra and
coalgebra/algebra.\footnote{Note that we take the ordinary
pairing here and not the type of pairing usually employed in braided
categories. In particular, note that $B$ and $D$ do not live in the
same braided category.}
Furthermore, we demand the compatibility
condition of action and coaction
\[
 \langle h\act d, b\rangle 
 = \langle h,b\i1 \rangle \langle d, b\iu2\rangle \quad
 \forall h\in H, b\in B, d\in D .
\]
Then the bosonizations $D\rbos H$
and $B\rbos A$ are naturally dually paired via
\[
 \langle d\tens h, b\tens a\rangle\defeq \langle d,b\rangle
 \langle h,a\rangle \quad \forall h\in H, a\in A, b\in B, d\in D .
\]
\end{lem}

As an example of how the bosonization construction reduces to an
ordinary semidirect product for groups and Lie algebras we consider
the Poincar\'e group and its Lie and (enveloping) algebra as well as
their pairing.

\begin{ex}
\label{ex:poinc}
In the context of Example~\ref{ex:clor}
consider the Hopf algebra $\mink$ of (polynomial) functions on the
translation group of Minkowski space. It is generated by $\{x^\mu\}$
as a free commutative algebra with coproduct, antipode and conjugation
\[
 \cop x^\mu=x^\mu\tens 1+ 1\tens x^\mu,\quad
 \antip x^\mu= -x^\mu,\quad \overline{x^\mu}=x^\mu.
\]
It is a
(trivially braided) $\qg{Spin(3,1)}$-comodule Hopf algebra via the
coaction $x^\mu\mapsto \sum_\nu \Lambda^{\mu \nu}\tens x^\nu$.

The Hopf algebra of functions $\poinc$ on the covered Poincar\'e group
is
the bosonization $\mink\rbos\qg{Spin(3,1)}$. As an algebra it is the
commutative algebra generated by $\qg{Spin(3,1)}$ and $\mink$. The
coalgebra structure and antipode for $t_{i j}$ is that of
$\qg{Spin(3,1)}$. The coproduct and antipode for $x^\mu$ are given by
\[
 \cop x^\mu = x^\mu\tens 1 + \sum_\nu \Lambda^{\mu \nu}\tens
x^\nu, \quad \antip x^\mu=-\sum_\nu (\antip\Lambda^{\mu \nu}) x^\nu .
\]
Note that the
sub-Hopf algebra generated by $\{x^\mu,\Lambda^{\mu \nu}\}$ is the
uncovered Poincar\'e group $\mink\rbos\qg{SO(3,1)}$.
\end{ex}

\begin{ex}
\label{ex:lpoinc}
In the context of Example~\ref{ex:so31}
let $\mom$ denote the abelian Lie algebra of translation
generators in 4 dimensions with basis $\{P^\mu\}$ and real structure
$\overline{P^\mu}=P^\mu$. Its universal
enveloping Hopf algebra $\ealg(\mom)$ is a (trivially braided)
$\ealg(\lalg{so_{3,1}})$-module Hopf algebra by the action
\[
X\act P^\mu = \frac{1}{2}\sum_\nu \tr(\sigma^\nu \sigma(X)\sigma^\mu)
 P^\nu, \quad
\overline{X}\act P^\mu = \overline{X\act P^\mu}\quad\forall X\in
 \{H,E,F\} .
\]
The semidirect product of the Lie algebras is the Poincar\'e Lie
algebra. Correspondingly for the enveloping Hopf algebras
$\ealg(\lpoinc)=\ealg(\mom\rtimes
\llor)=\ealg(\mom)\rbos\ealg(\llor)$.

$\ealg(\mom)$ and $\mink$ are dually paired via $\langle P^\mu,x^\nu
\rangle=\delta^{\mu \nu}$. As this pairing is compatible with the
action of $\ealg(\llor)$ and coaction of $\qg{Spin}(3,1)$ in the sense
of Lemma~\ref{lem:pbos} it induces a pairing between the Hopf
algebras $\ealg(\lpoinc)$ and $\poinc$.
\end{ex}

\section{Reconstruction of Quantum Group Symmetry}
\label{sec:rec}

\begin{figure}
\begin{center}
\input{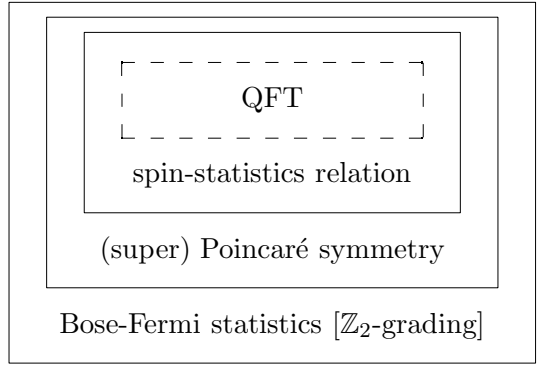}
\caption{The three layers of the reconstruction
of the quantum group symmetry. They correspond to successive
restrictions of categories.}
\label{fig:layers}
\end{center}
\end{figure}

\subsection{Poincar\'e Symmetry and Bose-Fermi statistics}
\label{sec:recpoinc}

Let us reconstruct the relevant braided monoidal category (and quantum
group) for a quantum field theory that is Poincar\'e symmetric,
has Bose-Fermi statistics, and obeys the spin-statistics theorem.
In fact, we simplify the discussion slightly by only considering the
$\SU2$ subgroup of $\Pc$
since it exhibits already all the relevant features.
We come back to the full Poincar\'e group at the end.
The construction proceeds in three ``layers'' corresponding 
respectively to the statistics, the symmetry group, and the
spin-statistics relation, see Figure~\ref{fig:layers}.

The first layer (the outermost box in Figure~\ref{fig:layers}) is an
underlying $\Z_2$-grading. More precisely, we
consider the braided category of $\Z_2$-graded vector spaces. This
distinguishes Fermions from Bosons and carries the Bose-Fermi
statistics. Reconstructing the quantum group that generates the
category, we obtain the function Hopf algebra $\CZ2$ of the group
$\Z_2$, but with the nontrivial cotriangular structure, described in
Example~\ref{ex:z2}.
The cotriangular structure provides the braiding
\[
 \psi(v\tens w)= (-1)^{|v|\cdot |w|} w\tens v ,
\]
which encodes Bose-Fermi statistics.

The second layer (the intermediate solid box in Figure~\ref{fig:layers})
is the symmetry group, in this case $\SU2$.
We wish to restrict our category of $\Z_2$-graded vector spaces
to those spaces that are representations of the group $\SU2$ as well.
Furthermore, we require the group action to respect
the grading. This means that we can view $\SU2$ as living
in the $\Z_2$-graded category itself. More precisely, taking the quantum
group point of view, the corresponding function Hopf algebra $\CSU2$
is an object in the category. It is purely even under the
$\Z_2$-grading. The braided category which encodes both the Bose-Fermi
statistics as well as the $\SU2$-symmetry is then the subcategory of
comodules of $\CSU2$ \emph{inside} the category of $\Z_2$-graded
vector spaces. However, according to the reconstruction theorem we can
express the braided category as a category of representations of
just one quantum group. We are here precisely in the situation of
Theorem~\ref{thm:cqtrbos} and
Proposition~\ref{prop:ctrbos} which tell us that this quantum group
is obtained from $\CZ2$ and $\CSU2$ by a kind of semidirect product,
called \emph{bosonization} $\CSU2\rbos\CZ2$. In the case at hand this
reduces just to the ordinary tensor product $\CSU2\tens\CZ2$
since $\CSU2$ is purely even, i.e., trivial as a representation
of $\CZ2$.
Using the basis of $\CSU2$ given in Example~\ref{ex:csu2} the tensor
product $\CSU2\tens\CZ2$ has a basis
$\{t^l_{m\,n}, t^{l}_{m\,n}g\}$.
Its coquasitriangular structure
is given by
\[
\cR(t^l_{m\,n}g^k\tens t^{l'}_{m'\,n'}g^{k'}) =
 (-1)^{k k'}\delta_{m\,n} \delta_{m'\,n'} .
\]

The third and final layer (the innermost solid box in
Figure~\ref{fig:layers})
consists of removing those representations
that have the wrong spin-statistics relation.
We only allow representations where either the spin-label is
integer and the $\Z_2$-degree even or the spin-label non-integer and the
$\Z_2$-degree odd. The coaction for a spin-$l$ representation thus
must take the form
\[
 v_m\mapsto \sum_n t^l_{m\, n} g^{2l}\tens v_n .
\]
Allowing only representations of this form
is equivalent to restricting
the Hopf algebra $\CSU2\tens\CZ2$ to the sub-Hopf algebra
spanned by $\{t^l_{m\,n}g^{2l}\}$. More generally, restricting a
monoidal category to a monoidal subcategory corresponds by
Tannaka-Krein reconstruction exactly to restricting a Hopf algebra
to a sub-Hopf algebra.

By renaming the element $t^l_{m\,n}g^{2l}$ with $t^l_{m\,n}$ we
recognize that the Hopf algebra we arrive at is nothing but
$\CSU2$ again. However, the coquasitriangular structure we obtain
by restriction from $\CSU2\tens\CZ2$
is not the canonical (trivial) coquasitriangular
structure of $\CSU2$. Instead, it is given by
\begin{equation}
\cR(t^l_{m\,n}\tens t^{l'}_{m'\,n'}) =
 (-1)^{4 l l'}\delta_{m\,n} \delta_{m'\,n'} .
\label{eq:ctrsu2}
\end{equation}
To make this distinction clear we denote the new coquasitriangular
Hopf algebra by $\CgSU2$.
It is precisely the one that was already found in \cite{Oe:spinstat}
by considering spin and statistics symmetries directly.

The construction generalizes to double covers of
space-time symmetry groups. Thus, assume a given space-time symmetry
group $G$ whose universal cover (i.e., quantum mechanically relevant
symmetry) is a double cover $\hat{G}$. Then, the
function Hopf algebra decomposes into a direct sum $\qg{\hat{G}}=
\qg{\hat{G}}_s\oplus \qg{\hat{G}}_a$ with $\qg{\hat{G}}_s=\qg{G}$
of functions that are symmetric
respectively antisymmetric with respect to interchange of the
sheets of the
cover. As the direct sum is a direct sum of coalgebras this
introduces a grading on the representations which corresponds
precisely to
spin (i.e., grading in integer versus half-integer spin).
Assuming the usual spin-statistics relation we obtain as above
$\qg{\hat{G}}$ itself as the relevant symmetry quantum group but with
the cotriangular structure of the following lemma.
This generalizes (\ref{eq:ctrsu2}). See also \cite{Oe:spinstat}.

\begin{lem}
\label{lem:ctrdcov}
Let $H$ be a commutative Hopf algebra which is $\Z_2$-graded as an
algebra into a direct sum of subcoalgebras $H=H_0\oplus H_1$.
Then
\begin{equation}
 \cR(f\tens g)=(-1)^{|f| |g|}\cou(f)\cou(g) ,
\label{eq:ctrdcov}
\end{equation}
where $|f|$ is the degree of $f$ with respect to the grading,
defines a cotriangular structure on $H$.
\end{lem}

In particular, for the (covering) Lorentz group $\qg{Spin}(3,1)$ and
the full Poincar\'e group $\poinc$ the grading is
given by $|t^l_{i j}|=|\overline{t^l_{i j}}|=2l$  (mod 2) and
$|x^\mu|=0$. We denote
the versions with
the cotriangular structure (\ref{eq:ctrdcov}) by $\qg{Spin}'(3,1)$ and
$\poinc'$ respectively.

\begin{rem}
There did not seem to be
any intrinsic reason to put the $\Z_2$-grading encoding the Bose-Fermi
statistics ``below'' the conventional symmetry group.
The bosonization appearing above really is just an ordinary tensor
product. However, when we go to supersymmetric groups such as the
super-Poincar\'e group this is no longer true. In this case the
reconstruction really requires the $\Z_2$-grading to lie ``below'' as
now the group is non-trivially graded. See Sections~\ref{sec:frec}
and \ref{sec:recext}.
\end{rem}

\subsection{The Dual Context}
\label{sec:dualrec}

As it is more familiar to physicists we also describe the dual context
with Lie algebras and universal enveloping algebras. Thus, let
$\lalg{g}$ be the Lie group of a space-time symmetry group $G$ as
considered above. Its
universal enveloping Hopf algebra $\ealg(\lalg{g})$ is dually paired
with the function Hopf algebra $\qg{G}$ as well as with the function
Hopf algebra of the double cover $\qg{\hat{G}}$. We can describe
comodules of $\qg{\hat{G}}$ alternatively as modules of
$\ealg(\lalg{g})$. However, the global information about the
difference between $G$ and $\hat{G}$ that contains the information
about (integer versus half-integer) spin is lost. But we can recover
the information by adjoining an
element $\xi$ to $\ealg(\lalg{g})$ with $\xi^2=1$ which commutes
with all other
elements and has coproduct $\cop \xi=\xi\tens\xi$. We extend the
pairing with $\qg{\hat{G}}$ by defining
\[
\langle \xi, f\rangle = (-1)^{|f|} \cou(f)\quad\forall f\in
\qg{\hat{G}} .
\]
The action of $\xi$ on a representation should yield the eigenvalues
$1,-1$ depending on whether spin is ``integer'' or ``half-integer''. To
ensure this we need to formally identify $\xi$ with $(-1)^C$ where $C$
is a suitable operator having even/odd
eigenvalues on ``integer''/``half-integer'' spin representations.
(Note that given such an operator, $(-1)^C$ is automatically central,
idempotent, and group-like.)

Following the construction above then leads to this version of
$\ealg(\lalg{g})$ as the spin-statistics reduced (dual) quantum group
with the non-trivial triangular structure given by the following
lemma.

\begin{lem}
\label{lem:trcov}
Let $A$ be a cocommutative Hopf algebra with a central element $\xi$
satisfying
$\xi^2=1$ and $\cop\xi=\xi\tens\xi$.
Then it admits a triangular structure
\begin{equation}
\mR = \frac{1}{2}(1\tens 1+1\tens\xi+\xi\tens 1-\xi\tens\xi) .
\label{eq:trcov}
\end{equation}
\end{lem}

Thus, we see that we can do everything in the dual enveloping algebra
context as well, though at the price that the global structure needs
to be added by hand. This is one reason why we prefer the
function algebra setting.

In fact, we could have performed the reconstruction from the beginning
in the enveloping algebra setting. Then, the element $\xi$
(corresponding to the element $g$ generating $C'(\Z_2)$) would have
come from the (dual) bosonization construction for the enveloping
algebra. The final step of the spin-statistics reduction then
precisely corresponds to identifying $\xi=(-1)^C$.

\begin{ex}
\label{ex:lsu2p}
In the context of Example~\ref{ex:usl2}
we adjoin the element $\xi$ to $\ealg(\lalg{su_2})$
as described above which we formally equate with
$(-1)^C$ where $C\defeq 4EF+2H+H^2$. Since
$\langle C, t^l_{m n}\rangle =4 l(l+1)\delta_{m,n}$ we get
\[
 \langle \xi, t^l_{m n}\rangle =
 (-1)^{4l(l+1)}\delta_{m,n}=(-1)^{2l}\delta_{m,n}
\]
as required. We denote this version of the enveloping algebra with the
triangular structure (\ref{eq:trcov}) by $\ealg'(\lalg{su_2})$.

We proceed similarly for Example~\ref{ex:so31} and define
$\ealg'(\lalg{so_{3,1}})$ with the operator
$C+\overline{C}$ where $C$ is as defined above and
equate $\xi= (-1)^{C+\overline{C}}$.

We define the bosonization 
$\ealg'(\lpoinc)=\ealg(\mom)\rbos\ealg'(\llor)$ analogous to
Example~\ref{ex:lpoinc} where $\xi$ acts
trivially on $P^\mu$.
\end{ex}

\subsection{Formalized Reconstruction}
\label{sec:frec}

We now formalize and generalize the procedure of reconstructing
the symmetry quantum group, exposing more clearly the role of the
different layers. (This section is somewhat more technical and
can be omitted by non-experts.)

The first layer is as before the underlying statistics.
We generalize it from a $\Z_2$-grading given by $\CZ2$ to an arbitrary
cotriangular Hopf algebra $H$. That is, the statistics is now
encoded by the category of (left) $H$-comodules $\catmodcl{H}$.
Cotriangularity implies that the braiding is symmetric, i.e., the
braiding and its inverse are identical.
This is in fact the limit of validity of the traditional separation of
spin and statistics: When the braiding is non-symmetric such a
separation is no longer possible. This is essentially because
Proposition~\ref{prop:ctrbos} does not generalize to the
coquasitriangular case. To put it differently: The bosonization $B\rbos
H$ admits an induced coquasitriangular structure from $H$ in general only
if $H$ is cotriangular.

For the second layer, the symmetry, we require now a braided
commutative Hopf algebra $B$
in $\catmodcl{H}$. This generalizes
the concepts of group and supergroup to arbitrary braiding.
Again, using Theorem~\ref{thm:cqtrbos} and
Proposition~\ref{prop:ctrbos}, the quantum group that generates
the braided category of representations of $B$ inside $\catmodcl{H}$
is the bosonization $B\rbos H$.

For the third layer, a spin-statistics relation obviously
requires that we have a ``spin'' that we can put in correspondence with
the statistics. In the previous section that came from the group
$\SU2$. In fact, the only relevant part of it (integer or
non-integer spin) is encoded
in the subgroup $\Z_2$ of $\SU2$. More generally, we need the same
(quantum) group as the one encoding the statistics but now as a quotient
(``subgroup'' in group language) of $B$. In other words, we require
a surjection of braided Hopf algebras $\sigma:B\tos H$ in
$\catmodcl{H}$, where $H$ is trivial as an $H$-comodule.

To impose now the spin-statistics relation we observe that the surjection
$\sigma:B\tos H$ gives rise to a surjection of
cotriangular Hopf algebras
\[
 \tilde{\sigma}:\Bt\defeq B\rbos H\tos H\rbos H
\]
upon bosonization.
In fact, $H\rbos H=H\tens H$ since the coaction of $H$ on $H$ was
taken to be trivial. For an arbitrary object $V$ in the category of
$\Bt$-comodules, spin and statistics are given by the induced
coaction of $H\tens H$. Denoting the coaction by $\beta:V\to \Bt\tens
V$ this induced coaction is given by
$\tilde{\beta}\defeq(\tilde{\sigma}\tens\id)\circ\beta:
 V\to H\tens H\tens V$.
To understand what it means to satisfy the spin-statistics relation
let us think for a moment in the more familiar language of groups.
Thus, let us think that (the dual of) $\tilde{\beta}$ defines an
action of two copies of the spin-statistics group $G$ on $V$. Now, $V$
obeys the spin-statistics relation if it is in the same representation
with respect to both copies of $G$. We can express this formally by
saying that the action of $G\times G$ factors through the action of a
single copy of $G$ by the group multiplication $G\times G\to G$.
Translating this back to the quantum group language means that the
image of $\tilde{\beta}$ must lie in $(\cop H)\tens V$ where $\cop H$
is the image of $H$ in $H\tens H$ under the coproduct.
This is precisely ensured by restricting $\Bt$ to the largest sub-Hopf
algebra
$A\subseteq \Bt$ so that $\tilde{\sigma}(A)\subseteq \cop H$. In fact,
$A$ is not in general the preimage $\tilde{\sigma}^{-1}(\cop H)$ as
this is not necessarily a Hopf algebra.
One can derive a stronger condition directly from the properties of a
comodule $V$.
In fact, it is not enough that $V$ satisfies the
spin-statistics relation in the form $\tilde{\beta}(V)\subseteq (\cop
H)\tens V$. But applying the coproduct several times (or alternatively
the coproduct on $H\tens H$) the corresponding condition must hold for
any copy of $H\tens H$ that appears in the image. This leads to
the condition $(\id\tens\tilde{\sigma}\tens\id)\circ \cop^2(A)
\subseteq A\tens\cop H\tens A$ which defines a bialgebra. As we
require a Hopf algebra we need to impose the even more restrictive
condition $(\id\tens\tilde{\sigma}\tens\id)\circ \cop^2(A)
\subseteq A\tens (\cop H\cap\tau(\cop H))\tens A$ where
$\tau$ is the transposition map. This corresponds to requiring for a
module $V$ that also its dual satisfies the spin-statistics relation.
We call $A$ the \emph{spin-statistics reduction} of $\Bt$
and formalize as follows:

\begin{dfn}
\label{def:ssred}
Let $H$ be a cotriangular Hopf algebra.
Let $\sigma:B\tos H$ be a solution of the triangular group extension
problem in $\catmodcl{H}$, where $H$ is equipped with the trivial
comodule structure under itself.
Consider the induced map $\tilde{\sigma}:\tilde{B}\defeq B\rbos H
\tos H\rbos H=H\tens H$. Then $A\subseteq \Bt$ defined as the subspace
satisfying
\begin{equation}
 (\id\tens\tilde{\sigma}\tens\id)\circ \cop^2(A)
\subseteq A\tens (\cop H\cap\tau(\cop H))\tens A
\label{eq:ssred}
\end{equation}
is a sub-Hopf algebra 
called the \emph{spin-statistics reduction}.
\end{dfn}

\subsection{Extensions}
\label{sec:recext}

We now turn to the question of whether and how a group extension in
the conventional (or triangular) sense gives rise to a quantum group
extension.

Let us consider the Bose-Fermi case first. Thus, we have a
$\Z_2$-graded Hopf algebra $B$ (e.g., the ordinary Poincar\'e group
which is just trivially graded) and a $\Z_2$-graded extension $C$ of it
(e.g., the super-Poincar\'e group). That is, we have a $\Z_2$-graded
group extension $C\tos B$ in $\catmodcl{\CZ2}$.
In general, we have some cotriangular Hopf algebra $H$ in place of
$\CZ2$ and $\rho:C\tos B$ is a triangular extension in the
sense of Definition~\ref{def:gext}.
On the other hand, both $B$ and $C$ are also
solutions to the extension problem for the ``spin'' $H$ as a
trivial comodule in $\catmodcl{H}$. Thus we have surjections
$\sigma_B:B\tos H$ and $\sigma_C:C\tos H$ as well and
$\sigma_C=\sigma_B\circ\rho$. The categorial equivalence
of Proposition~\ref{prop:ctrbos} lifts these to surjections of
cotriangular Hopf algebras
$C\rbos H\tos B\rbos H\tos H\rbos H=H\tens H$.

We now apply the spin-statistics reduction of Definition~\ref{def:ssred}.
Denote the reduced quantum groups by
$B'$ and $C'$. The image of $C'$ obviously satisfies the reduction
condition itself 
and thus we have a map $C'\to B'$ as the restriction of
$C\rbos H\tos B\rbos H$. However, this map is not necessarily
surjective. Thus, we do not necessarily obtain a solution of the
quantum group extension problem, but something weaker.
The triangular extension could behave ``badly'' with respect to the
spin-statistics relation.

\section{Nothing beyond Supersymmetry}
\label{sec:nobeyond}

In this section we perform in a sense the opposite operation
to the reconstruction of Section~\ref{sec:recext}.
We show, for the case of Bose-Fermi statistics,
that any solution of the quantum group extension problem
can be induced from a solution of the triangular group extension
problem.
Returning to the initial question wether a unification of external and
internal degrees of freedom beyond supersymmetry is possible, this
implies a negative answer. More precisely, any extension of the
symmetry quantum
group of ordinary quantum field theory (in at least 3 spatial
dimensions) can already be obtained through the known
(supersymmetric) ones.

We first consider the special case where the quantum group to be extended
is just the spin (and thus also statistics) generating one.
This is our main theorem.

\begin{thm}
\label{thm:qexttogext}
Let $\pi:A\tos \CZ2$ be a solution of the quantum group
extension problem for $\CZ2$. Then, there is a solution
$\sigma:B\tos \CtZ2$ of the $\Z_2$-graded group extension problem
in $\catmodcl{\CZ2}$ where $\CZ2$ coacts trivially on itself,
such that $A$ is the spin-statistics reduction
of $B\rbos \CZ2$.
\end{thm}
\begin{proof}
Define the space $\Bt\defeq A\tens \CZ2$ and a surjection
$\sigmat:\Bt\to \CtZ2\tens \CZ2$ given by
$a\tens h\mapsto \pi(a\i1) \tens \pi(a\i2) h$. This gives rise to the
sequence
\begin{equation}
 \Bt\xrightarrow{\sigmat} \CtZ2\tens \CZ2
  \xrightarrow{\cou\tens\id} \CZ2 .
\label{eq:constseq}
\end{equation}
We give $\Bt$ the tensor product
coproduct structure, the subalgebra structures of $A$ and $\CZ2$ and
the cross relations induced by the pull-back of the cotriangular
structure of $\CZ2$. This makes (\ref{eq:constseq}) into a sequence of
cotriangular Hopf algebras. (Note that the coquasitriangular structure on
$\CtZ2\tens \CZ2$ is thus trivial on the first component, hence the
notation without the prime.) Now consider the injection of
cotriangular Hopf algebras
$i:\CZ2\to\Bt$ given by $h\mapsto 1\tens h$.
The surjection $\Bt\to \CZ2:(\cou\tens\id)\circ\sigmat$ inverts $i$.
Thus, according to Theorem~\ref{thm:cqtrbos} there is a braided Hopf
algebra $B$
in the category of left $\CZ2$-comodules so that $\Bt=B\rbos \CZ2$.
We can recover $B$ (as an algebra) as $^{\CZ2} \Bt$ and observe that
on this space the map $\sigmat$ restricts to the subspace $\CtZ2\tens 1$
of $\CtZ2\tens \CZ2$. Identifying $\CtZ2\tens 1$ as $^{\CZ2}
(\CtZ2\tens \CZ2)$ we obtain precisely a surjection $\sigma:
B\tos\CtZ2$ as required.

It remains to show that $A$ is the spin-statistics reduction $A'$ of
$\Bt$. For this observe that th condition (\ref{eq:ssred}) of
Definition~\ref{def:ssred} implies
$(\id\tens\tilde{\sigma}\tens\id)\circ \cop^2(A')
\subseteq \Bt\tens (\cop\CZ2)\tens \Bt$. This in turn implies for an
element $a\tens 1+
b\tens g$ in $A'$ that
$(a\i1\tens 1)\tens 1 \tens (a\i2\tens 1)
+ (b\i1\tens g)\tens g\tens (b\i2\tens g)\in \Bt\tens 1\tens \Bt$ by
composition with $\id\tens(\cdot\circ(\antip\tens\id))\tens\id$. Thus
$b=0$ and it follows that $A'\subseteq A$. On the other hand clearly
$A\subseteq A'$ and thus $A=A'$. This completes the proof.
\end{proof}

The case of a general extension is obtained by considering two
extensions of the spin generating group $\CZ2$ with a surjection and then
observing that the surjection survives the transition from
the quantum group context to the $\Z_2$-graded group context.

\begin{prop}
\label{prop:qexttogext}
In the context of Theorem~\ref{thm:qexttogext} consider two solutions
of the quantum group extension problem with a surjection
$A'\tos A\tos \CZ2$. Then there is an induced surjection between
the corresponding $\Z_2$-graded groups $B'\tos B\tos H$.
Thus, to the quantum group extension $A'\tos A$ corresponds the
$\Z_2$-graded group extension $B'\tos B$.
\end{prop}

In the case of ordinary quantum field theory the symmetry to
be extended is the Poincar\'e group $B=\qg{\Pc}$ and it corresponds
to the quantum group $A=\qg{\Pc}'$ introduced in
Section~\ref{sec:recpoinc} (i.e., the Poincar\'e group with the
cotriangular structure (\ref{eq:ctrdcov})). The statement of the
Proposition~\ref{prop:qexttogext} is now that any solution
$A'\tos \qg{\Pc}'$ of the quantum
group extension problem is induced from a
solution $B'\tos \qg{\Pc}'$ of the $\Z_2$-graded group
extension problem.
But this is
nothing but the supergroup extension problem described in
the introduction.
Thus, for ordinary QFT all the solutions of the
quantum group extension problem are induced from solutions of the
supergroup extension problem.
The analysis of \cite{HaLoSo:allsusysmat} remains exhaustive in our
generalized framework.

\section{Examples and Applications}
\label{sec:exappl}

In this section we wish to demonstrate the usefulness of quantum
geometric methods to supersymmetry.
Notions of homogeneous space, quantum principal bundle, exterior
derivative, all generalize from ordinary geometry to quantum
geometry. In particular they apply to supergroups,
superspaces, super-derivatives etc. Constructions in
quantum geometry are just as easy for ``super''-objects as they are
for ordinary objects. Furthermore, they generalize far beyond the
$\Z_2$-graded case (although we shall not consider this here), see in
particular \cite{Maj:qgroups}. For example they
apply to more general anyonic and braid statistics, see
\cite{Oe:spinstat}. An application to fractional
supersymmetry which fits into this framework is
\cite{DMAP:geofsusy,Dun:braidfsusy}.

\subsection{Elements of Quantum Geometry}

We start by introducing the basic notions and give elementary examples
(from ordinary geometry). This section is mostly text book
knowledge, see \cite{Maj:qgroups, KlSc:qgroups}. For quantum principal
bundles see \cite{BrMa:qgauge}, although our version here is more
elementary. For statements from ordinary differential geometry, see e.g.\
\cite{KoNo:diffgeo1}. 

\begin{dfn}
Let $H$ be a Hopf algebra, $\beta:P\to P\tens H$ a right $H$-comodule
algebra. Set $B\defeq P^H=\{p\in P|\beta(p)=p\tens 1\}$ and define
$\chi: P\tens P\to P\tens H$ by
$\chi=(\cdot\tens\id)\circ (\id\tens\beta)$. If $\chi$ is surjective
we call the triple $(P,B,H)$ a \emph{quantum principal bundle}.
\end{dfn}

To see how this definition reduces to the usual one for ordinary
principal bundles consider a group $G$ acting on a manifold $E$. This
gives rise to a coaction $\beta:\falg(E)\to
\falg(E)\tens\falg(G)$. The surjectivity of $\chi$ then means
precisely that the map $G\times E\to E\times E$ defined by
$(g,p)\mapsto (g p, p)$ is injective, i.e., that $G$ acts freely and
thus defines a principal bundle.\footnote{In fact, it is also
necessary that $G$ acts properly. This is a somewhat
technical condition, strongly related to the class of functions we
consider. It is satisfied for ``good'' cases such as when $G$ is
compact.
We thus consider this condition as ``not being visible'' in our
algebraic setting.}
The base space $M$ is the space of orbits
of $G$ in $E$ and $\falg(M)$ is precisely $\falg(E)^{\falg(G)}$.

\begin{dfn}
Let $\pi:A\tos H$ be a surjection of Hopf algebras. $A$ is a right
$H$-comodule algebra by $\beta_R=(\id\tens\pi)\circ\cop$ and a left
$H$-comodule algebra by $\beta_L=(\pi\tens\id)\circ\cop$.
The space
$A^H\defeq\{a\in A|\beta_R(a)=a\tens 1\}$ forms a left $A$-comodule
algebra via the coproduct of $A$ (and thus also a left $H$-comodule
algebra via $\beta_L$). It is called a
\emph{quantum homogeneous space}.
\end{dfn}

Again we consider how this definition reduces to the one for ordinary
homogeneous spaces. Let $G\toi K$ be an injection of groups. This
gives rise to a surjection of Hopf algebras
$\pi:\falg(K)\tos\falg(G)$. The functions on the homogeneous space
$K/G$ by the induced action of $G$ on $K$ are precisely the functions
on $K$ invariant under this group action, i.e.,
$\falg(K/G)\cong\falg(K)^{\falg(G)}$. In particular, a homogeneous
space gives rise to a principal bundle.

\begin{rem}
\label{rem:qhomqpb}
Given a Hopf algebra surjection $A\tos H$ we obtain a
quantum principal bundle $(A,A^H,H)$. Furthermore, this bundle is left
$A$-equivariant, i.e., $A$ carries a left coaction by $A$ itself which
commutes with the right coaction of $H$ and thus descends to $A^H$.
\end{rem}

This is in exact analogy to the situation in ordinary geometry.

\begin{rem}
\label{rem:bosqhom}
Given a coquasitriangular Hopf algebra $H$ and a braided Hopf algebra
in $\catmodcl{H}$ the bosonization gives rise to a Hopf algebra
surjection $B\rbos H\tos H$ (see Theorem~\ref{thm:cqtrbos}). In
particular, we obtain a quantum homogeneous space. In fact this
precisely recovers $B$ itself (as an algebra) $B=(B\rbos H)^H$.
\end{rem}

This generalizes the situation of a semidirect product of Lie groups
giving rise to a homogeneous space.

\begin{rem}
\label{rem:qpspin}
Note that for a homogeneous space $K/G$ the principal
bundle $(K,K/G,G)$ can be identified with (a reduction of) the frame
bundle on $K/G$. Furthermore, if $K/G$ is Riemannian such that $K$
consists of isometries, the bundle $(K,K/G,G)$ can be identified with
(a reduction of) the bundle of orthonormal frames on $K/G$.
Now, if $K/G$ is orientable the bundle decomposes into
two connected components (corresponding to the two possible orientations).
We take the one corresponding to the orientation preserving subgroup
of $K$ and denote it $(K',K/G,G')$.
$K/G$ admits a spin structure iff $K'$ admits a 
double cover $\hat{K}$. The spin bundle is then
$(\hat{K},K/G,\hat{G})$ where $\hat{G}$ is the corresponding double
cover of $G'$.
\end{rem}

\begin{ex}
\label{ex:sphere}
Consider $S^2$ as the homogeneous space $SO(3)/SO(2)=SU(2)/U(1)$.
The injection $U(1)\toi SU(2)$ becomes a surjection $\qg{SU}(2)\tos
\qg{U}(1)$ from the quantum group point of view. $\qg{U}(1)$ has a
(Fourier, i.e., Peter-Weyl) basis $\{g^m\}$ with $m\in\Z$,
coproduct $\cop g^m=g^m\tens g^m$, counit $\cou(g)=1$, and antipode
$\antip g^m=g^{-m}$. The Hopf algebra surjection is given by
\begin{equation}
 t^l_{m n}\mapsto g^{2m}\delta_{m n}
\label{eq:surj}
\end{equation}
in the basis of
Example~\ref{ex:csu2} for $\qg{SU}(2)$.

The surjection (\ref{eq:surj}) induces the right coaction
$t^l_{m n}\mapsto t^l_{m n}\tens g^{2n}$ of $\qg{U}(1)$ on
$\qg{SU}(2)$ leading to the algebra $\qg{S}^2$ as its
right invariant subspace. A basis of it is given by
$\{t^l_{n 0}\}$ with $l\in N_0$. These are precisely the spherical
harmonics on $S^2$.

Note that according to Remark~\ref{rem:qpspin} we can view
$(\qg{SU}(2),\qg{S}^2,\qg{U}(1))$ as the spin-bundle on $\qg{S}^2$.
\end{ex}

We now turn to the concept of quantum differentials that generalizes
the concepts of differential 1-forms and vector fields to the quantum
geometric realm.

\begin{dfn}[\cite{Wor:calculi}]
Let $A$ be a Hopf algebra. Let $\Omega$ be a bicovariant bimodule over
$A$. That is, a left and right $A$-module and a left and right
$A$-comodule, such that actions and coactions commute in the obvious
ways. Assume there is a bicomodule map $\xd:A\to\Omega$. That is,
$\xd$ is a left and right $A$-comodule map. If the Leibniz rule
\[
 \xd (a b)=\xd(a) b+ a\xd(b),\quad\forall a,b\in A 
\]
holds and the map $A\tens A\to\Omega$ given by $a\tens b\mapsto a\xd
b$ is surjective, then we call $\Omega$ a (bicovariant first-order)
\emph{differential calculus}.
\end{dfn}

\begin{prop}[\cite{Wor:calculi}, see also \cite{Maj:classcalc}]
Let $A$ and $H$ be Hopf algebras that are non-degenerately dually
paired. Then, differential calculi on $A$ correspond to subspaces of
$L\subseteq H'\defeq \ker\cou\subset H$ with the following properties:

(a) $L$ is invariant under the left coaction of $H$ defined by
$\eta\mapsto \eta\i1\tens \eta\i2-\eta\tens 1$.
That is, the coaction $H'\to 
H\tens H'$ defined in this way descends to $L\to H\tens L$.

(b) $L$ is invariant under the left action of $H$ given by
$h\act \eta = h\i1 \eta \antip h\i2$. That is, the action $H\tens H'\to H'$
defined in this way restricts to $H\tens L\to L$.

The space $L$ can be thought of as the space of right-invariant vector
fields that act on $A$ as ``derivatives'' from the right via
\[
 L\tens A\to A:\eta\tens a\mapsto
 \partial_\eta(a)\defeq\langle \eta,a\i1\rangle a\i2
\]
for $\eta\in L$ and $a\in A$.
Dually, $\Gamma\defeq L^*$ is the space
of right-invariant 1-forms.
$\Gamma$ carries a left action of $A$ determined by the
coaction (a) via
\[
 \langle \eta,a\act\omega\rangle =
  \langle \eta\i1,a \rangle \langle\eta\i2,\omega\rangle
 -\langle \eta,a \rangle \langle 1,\omega\rangle
\]
for $a\in A, \eta\in L, \omega\in\Gamma$.
$\Gamma$ carries a left coaction 
$\omega\mapsto\omega\ib1\tens \omega\ibu2$
of $A$ determined by the action (b)
via
\[
 \langle h,\omega\ib1\rangle\langle\eta,\omega\ibu2\rangle
 =\langle h\i1\eta\antip h\i2,\omega\rangle
\]
for $h\in H,\eta\in L,\omega\in\Gamma$.
The corresponding differential calculus $\Omega$ on $A$ is
isomorphic to $\Gamma\tens A$ as a vector space. Its right $A$-module and
comodule structure are given by multiplication and comultiplication on
$A$. Its left $A$-module and comodule structure are the tensor product
ones. The map $d$ is recovered from the derivative as 
$\xd(a)=\sum_i \omega_i \partial_{\eta_i}(a)$ with $\eta_i$ a basis of
$L$ and $\omega_i$ the dual basis of $\Gamma$.

The left coaction $\eta\mapsto \eta\ib1\tens\eta\ibu2$ of $A$ on $L$
that makes the derivative map $L\tens A\to A$ covariant is determined
through the action of condition (b) by
\[
\langle \antip h, \eta\ib1\rangle \eta\ibu2=h\i1 \eta \antip h\i2 .
\]
for $h\in H,\eta\in L$.
\end{prop}

\begin{rem}
Let $G$ be a Lie group and $\lalg{g}$ its Lie algebra. We set
$A=\falg(G)$ and $H=\ealg(\lalg{g})$ above and recover the usual
differentials with $L=\lalg{g}$. Note that the coaction of condition
(a) becomes trivial while the action of condition (b) is by the Lie
bracket.
\end{rem}

\begin{ex}
\label{ex:derivsu2}
Consider the dually paired quantum groups $\qg{SU}(2)$ and
$\ealg(\lalg{su_2})$ described in Example~\ref{ex:usl2}. The ordinary
tangent space is $\lalg{su_2}$ with basis $E,F,H$
leading to the derivatives
\begin{gather*}
 \partial_H(t^l_{m n})=2 m t^l_{m n},\quad
 \partial_E(t^l_{m n})=\sqrt{(l-m+1)(l+m)} t^l_{m-1,n},\\
 \partial_F(t^l_{m n})=\sqrt{(l+m+1)(l-m)} t^l_{m+1,n} .
\end{gather*}

Note that this result extends to the dually paired quantum groups
$\qg{Spin}(3,1)$ and $\ealg(\llor)$ precisely as in the transition
from Example~\ref{ex:usl2} to Example~\ref{ex:so31}.
\end{ex}

\begin{ex}
\label{ex:derivpoinc}
Consider the dually paired Poincar\'e group $\poinc$ and its Lie
algebra $\lpoinc$ of Examples~\ref{ex:poinc} and \ref{ex:lpoinc}.
The derivatives given by $\lpoinc$ on $\poinc$ come out as in
Example~\ref{ex:derivsu2} supplemented by
\begin{gather*}
 \partial_X(x^\mu)=
  \frac{1}{2}\sum_\nu\tr(\sigma^\mu\sigma(X)\sigma^\nu), \quad
 \partial_{\overline{X}}(x^\mu)=\overline{\partial_X(x^\mu)}, \quad
 \forall X\in\{H,E,F\}, \\
 \partial_{P^\mu}(t^l_{m n})
 = \partial_{P^\mu}(\overline{t^l_{m n}})=0,\quad
 \partial_{P^\mu}(x^\nu)=\delta^{\mu \nu} .
\end{gather*}
\end{ex}

\subsection{Semidirect Superextensions}

In this section we consider semidirect superextensions which are
simple examples of superextensions. These serve at the same time as a
preparation for the more involved superextensions considered later.

For our present mathematical purposes we give the following definition
of ``superextension'' as a minor modification of
Definition~\ref{def:qext}.

\begin{dfn}
\label{def:sext}
Let $H$ be a coquasitriangular Hopf algebra. Then $A$ is called a
\emph{finite non-trivial extension} or \emph{superextension} of $H$ if
(a) $A\tos H$ is a surjection of
coquasitriangular Hopf algebras, (b) there is no Hopf algebra $K$
such that $A\cong H\tens K$ as a Hopf algebra and, (c) $A^H$ is finite
dimensional.
\end{dfn}

\begin{rem}
In the quantum principal bundle picture $(A,A^H,H)$ condition (b)
corresponds to the bundle being non-trivial. Condition (c) says
that the algebra of functions on the base space is finite
dimensional or the base space itself ``zero-dimensional''.
Physically speaking it means that the number of
superfield components is finite.
\end{rem}

We shall be interested in the case where $H$ is a cotriangular Hopf
algebra of the type of Lemma~\ref{lem:ctrdcov} as this is the physically
interesting situation (see Section~\ref{sec:rec}). Dually, we shall
as well consider dual superextensions (defined in the obvious way)
for triangular Hopf algebras of the type of Lemma~\ref{lem:trcov}.

An important (well known) supergroup is the analog of the translation
group on $\R^n$ defined as follows.

\begin{lem}
\label{lem:sutra}
Let $\Theta_n$ be the unital algebra generated by
$\{\theta_1,\dots,\theta_n\}$ with relations
$\theta_i\theta_j=-\theta_j\theta_i$. It is $\Z_2$-graded by
$|1|=0$, $|\theta_i|=1$. It extends to a $\Z_2$-graded commutative
Hopf algebra by
the coalgebra structure $\cop\theta_i=1\tens \theta_i+\theta_i\tens
1$. Counit and antipode are given by $\cou(\theta_i)=0$ and
$\antip \theta_i=-\theta_i$.

Furthermore, $\Theta_n$ is self-dual via the pairing generated by
$\langle\theta_i,\theta_j\rangle=\delta_{ij}$.\footnote{Note that we
take the 
pairing here in the usual sense for Hopf algebras and not the usual
sense for braided Hopf algebras. That is, we require
$\langle \phi, v w\rangle=\langle \phi\i1,v\rangle \langle
\phi\i2,w\rangle$ etc., and not with $\phi\i1,\phi\i2$ interchanged.}
\end{lem}

We are now ready to consider semidirect superextensions.

\begin{prop}
\label{prop:semiext}
In the context of Lemma~\ref{lem:ctrdcov}
let $V$ be a finite dimensional comodule of $H$ such that for the
coaction $\beta:V\to H\tens V$ the condition $\beta(V)\subseteq
H_1\tens V$ holds. Then, $\Theta_V\defeq\Theta_n$ with
$\{\theta_1,\dots,\theta_n\}$ a basis of $V$ is a braided $H$-comodule
Hopf algebra.

Furthermore, the semidirect product (bosonization) $\Theta_V\rbos H$
is a superextension of $H$.
It is generated as an algebra by $H$ and $\theta_i$ with
relations $\theta_i \theta_j=-\theta_j \theta_i$ and
$h \theta_i=(-1)^{|h|} \theta_i h$. The coalgebra structure on the
$H$ is the given one and for $\theta_i$ we have
\[
 \cop \theta_i =\theta_i\tens 1+ \beta(\theta_i) .
\]
\end{prop}

\begin{cor}
Let $H$ be a commutative Hopf algebra generated by the $n$-dimensional
matrix coalgebra
$T$ and equipped with the cotriangular structure
\begin{equation}
 \cR(t_{ij}\tens t_{kl})=-\cou(t_{ij})\cou(t_{kl})
 =-\delta_{ij}\delta_{kl} .
\label{eq:ctrcov}
\end{equation}
Then, $\Theta_n$ is a (braided) $H$-comodule Hopf algebra.

Furthermore, the semidirect product (bosonization) $\Theta_n\rbos H$
is a superextension of $H$.
It is generated as an algebra by $t_{ij}$ and $\theta_i$ with
relations $\theta_i \theta_j=-\theta_j \theta_i$ and
$\theta_i t_{jk}=-t_{jk} \theta_i$. The coalgebra structure on the
$t_{ij}$ is the matrix coalgebra structure while for $\theta_i$ we
obtain
\[
 \cop \theta_i =\theta_i\tens 1+ \sum_j t_{i j}\tens \theta_j .
\]
\end{cor}

\begin{rem}
Note that while the above seems to be adapted to real matrix groups it
works equally well for complex matrix groups. In that case, double the
range of the indices and define $t_{i+n,j+n}\defeq \overline{t_{i,j}}$
and set $t_{i,j+n}=0=t_{i+n,j}$.
\end{rem}

We can equally consider the dual setting with the ``enveloping algebra''
counterpart of $H$, although we need to adjoin an extra generator in
this case as described in Section~\ref{sec:dualrec}.

\begin{prop}
In the context of Lemma~\ref{lem:trcov}
given an $n$-dimensional $A$-module $V$ with basis $\{Q_1,\dots,Q_n\}$
such that $\xi\act Q_i=-Q_i$, it extends to the (braided)
$A$-module Hopf algebra $\Theta_V\defeq \Theta_n$.

Furthermore, the semidirect product (bosonization) $\Theta_V\rbos A$
is a dual superextension of $A$.
It is generated as an algebra by $A$ and $\Theta_V$ with
cross-relations 
$a Q_i =(a\i1\act Q_i) a\i2\,\forall a\in A$.
For primitive elements of $A$ the latter take the commutator form
$[a,Q_i]=a\act Q_i$.
The coalgebra structure on
$A$ is the given one and for $Q_i$ we obtain
$\cop Q_i=Q_i\tens 1+\xi\tens Q_i$.
\end{prop}

\begin{ex}
\label{ex:extsu2}
Consider the superextension $\Theta_2\rbos \ealg'(\lalg{su_2})$ where
$\Theta_2$ is in the fundamental representation. Explicitly, we denote
a basis of $\Theta_2$ by $\{Q_\pm\}$ and define the action with
$\sigma(X)$ given as in Example~\ref{ex:so31} as
\begin{gather*}
 X\act Q_i = \sum_j \sigma(X)_{j i} Q_j\quad\forall X\in\{H,E,F\},
 \qquad\text{or explicitly},\\
 E\act Q_+=0,\quad E\act Q_-=Q_+,\quad
 F\act Q_+=Q_-,\quad F\act Q_-=0,\\
 H\act Q_\pm=\pm Q_\pm,\quad \xi\act Q_\pm=-Q_\pm .
\end{gather*}
The cross-relations are immediate from that:
\begin{gather*}
 [E,Q_+]=0,\quad [E,Q_-]=Q_+,\quad
 [F,Q_+]=Q_-,\quad [F,Q_-]=0,\\
 [H,Q_\pm]=\pm Q_\pm,\quad \xi Q_\pm=-Q_\pm\xi .
\end{gather*}
\end{ex}

\begin{ex}
The pairing of $\Theta_2\rbos \qg{SU}'(2)$ and
$\Theta_2\rbos\ealg'(\lalg{sl_2})$ (using Lemma~\ref{lem:pbos}) leads to
the natural quantum
tangent space with basis $E,F,H,Q_+,Q_-$. The corresponding
derivatives are the ones of Example~\ref{ex:derivsu2} extended by
\begin{gather*}
 \partial_E(\theta_+)=\theta_-,\quad
 \partial_E(\theta_-)=0,\quad
 \partial_F(\theta_+)=0,\quad
 \partial_F(\theta_-)=\theta_+,\\
 \partial_H(\theta_\pm)=\pm\theta_\pm,\quad
 \partial_{Q_\pm}(t^l_{m n})=0,\\
 \partial_{Q_+}(\theta_+)=1,\quad
 \partial_{Q_+}(\theta_-)=0,\quad
 \partial_{Q_-}(\theta_+)=0,\quad
 \partial_{Q_-}(\theta_-)=1
 \end{gather*}
\end{ex}

For homogeneous spaces we can easily obtain their
extensions to superspaces corresponding to superextensions of the
group. This is simply the corresponding quantum homogeneous
space.

\begin{ex}
\label{ex:semisphere}
Analogous to the ordinary 2-sphere as a homogeneous space of $SU(2)$
(Example~\ref{ex:sphere})
we can build a supersymmetric version as the quantum homogeneous
space of $\Theta_2\rbos \qg{SU}'(2)$ via the surjection $\pi:\Theta_2\rbos
\qg{SU}'(2)\tos \qg{U}(1)$. We call this the
\emph{semidirect super-sphere}. $\pi$ is simply given by the extension
map $\Theta_2\rbos\qg{SU}'(2)\tos \qg{SU}'(2)$ composed with the usual
surjection $\qg{SU}'(2)\tos\qg{U}'(1)$ given by (\ref{eq:surj}). Note
that this is a cotriangular Hopf algebra map upon choosing
$\cR(g^m\tens g^n)=(-1)^{m n}\delta_{m n}$ on $\qg{U}'(1)$
(hence the prime in the notation) corresponding to $U(1)$ covering
itself twice. Now the semidirect super-sphere is simply the subalgebra
$\qg{S}^2_2$
of $\Theta_2\rbos \qg{SU}'(2)$ which is (right) invariant under the
coaction induced by $\pi$. That is, it is the subalgebra with basis 
$\{t^l_{i,0},t^l_{i, -}\theta_{+},
t^l_{i, +}\theta_{-},t^l_{i,0}\theta_{+}\theta_{-}\}$.

Note that $\qg{S}^2_2$ is a left $\Theta_2\rbos \qg{SU}'(2)$-comodule
algebra (via the coproduct) by construction. It gives rise to a quantum
principal bundle $(\Theta_2\rbos \qg{SU}'(2), \qg{S}^2_2,
\qg{U}'(1))$ (see Remark~\ref{rem:qhomqpb}).
Upon ``reducing the base space'' to $\qg{S}^2$ it becomes the
spin-bundle $(\qg{SU}'(2),\qg{S}^2,\qg{U}'(1))$ of $\qg{S}^2$.
Thus, we can view it as the spin-bundle of $\qg{S}^2_2$.
\end{ex}

\subsection{Matrix Supergroups}

We now consider more complicated superextensions which are
super-analogues of matrix groups. These are well known in the theory
of supergroups, see e.g.\ \cite{Wit:supermf}. (However, our setting is
closer in spirit to \cite{Kos:gradmf}.)
Much of their treatment here is
along the lines of \cite{Man:qdefsuper} and \cite{Maj:qgroups} (where
even more general braidings are considered).
However, the ``physical'' quantum group versions of supergroups
(motivated from Section~\ref{sec:rec}) seem not to have
been considered previously.
In particular,
the quantum group version $\qg{M}'(m|n)$ as well as the treatment of the
$OSp$-supergroups in this context appear to be new.

Let $m,n$ be natural numbers.
Consider a $\Z_2$-graded vector space $V$ with basis $\{v_i\}$ for
$i\in\{1,\dots, m+n\}$ such that $|v_i|\defeq |i|$ with
\[
|i|\defeq\begin{cases} 1 &\text{if } i\in \{1,\dots, m\}\\
  0 &\text{if } i\in\{m+1,\dots, m+n\}\end{cases} .
\]
Assume further that a graded commutative Hopf algebra
(i.e., a supergroup) $H$
coacts (graded) from the left on
$V$. Explicitly, $v_i\mapsto \sum_j u_{i j}\tens v_j$.
As the coaction
is graded the grading on the elements $\{u_{i j}\}$ must be given by
$|u_{i j}|=|i|+|j|$.

Conversely, in order to construct such a graded commutative Hopf
algebra, we start (exactly as we would do in the theory of matrix
groups) with the universal graded commutative bialgebra that coacts on
$V$.

\begin{dfn}
Consider the matrix coalgebra generated
by $\{u_{i j}\}$ with $i,j\in\{1,\dots,m+n\}$. It becomes a
$\Z_2$-graded coalgebra (i.e., coproduct and counit respect the
grading) by defining $|u_{i j}|\defeq |i|+|j|$ (mod 2 understood).
Next, consider its tensor algebra and
extend the coproduct to it as a graded algebra map. We obtain a graded
bialgebra. Finally, we quotient by the graded commutativity relation
\[
u_{i j} u_{k l}=(-1)^{(|i|+|j|) (|k|+|l|)} u_{k l} u_{i j} .
\]
As this is
compatible with the coproduct we obtain a graded commutative
bialgebra $\qg{M}(m|n)$. We call it the \emph{matrix super-bialgebra} of
rank $(m|n)$.

Note that we can quotient by the relations $u_{i j}=0$ for $|i|\neq
|j|$ to obtain a tensor product of purely even commutative matrix
bialgebras
\begin{equation}
 \qg{M}(m|n)\tos \qg{M}(m)\tens\qg{M}(n) .
\label{eq:commquot}
\end{equation}
\end{dfn}

In the above context of a graded commutative Hopf algebra coacting on
a graded vector space $V$ we consider
the dual space $V^*$ with a pairing $V\tens V^*\to\C$. It
naturally becomes a graded $H$-comodule by the coaction
$v^*_i\mapsto \sum_j (-1)^{|j|\cdot (|j|+|i|)} \antip u_{j i}\tens
v^*_j$ which leaves the pairing invariant (with $\{v^*_i\}$ the dual basis
to $\{v_i\}$).

Assume now that $V$ and $V^*$ are isomorphic as graded $H$-comodules
via a map $\eta:V^*\to V$ with $v^*_i\mapsto \sum_j\eta_{i j} v_j$. As
$\eta$ is bijective the inverse matrix $\eta^{-1}_{i j}$ exists
and $\eta_{i j}=\eta^{-1}_{i j}=0$ for $|i|\neq |j|$ as it is graded.
The fact that $\eta$ is a comodule map implies that the elements
$\antip u_{i j}$ of $H$ can be written in terms of the $u_{i j}$ as
\begin{equation}
\antip u_{i j}=\sum_{k,l}
  (-1)^{|l|\cdot (|l|+|j|)} \eta_{j k} u_{k l} \eta^{-1}_{l i} .
 \label{eq:invantip}
\end{equation}
This in turn implies the relations
\begin{gather}
\eta_{k l} \one = \sum_{i,j} (-1)^{|l|\cdot (|l|+|i|)}
 \eta_{i j} u_{j l} u_{i k} ,\label{eq:invrel1}\\
\eta^{-1}_{k l} \one = \sum_{i,j} (-1)^{|i|\cdot(|i|+|k|)}
 u_{l j} u_{k i} \eta^{-1}_{i j}
 \label{eq:invrel2}
\end{gather}
by the defining property of the antipode.

Conversely, we can construct a graded Hopf algebra by demanding it to
be the universal graded Hopf algebra with coaction on $V$ such that
$\eta$ is a graded comodule map. This is analogous to constructing 
a matrix group that leaves a non-degenerate bilinear form on its
defining representation (which can be seen as an isomorphism to the
dual representation) invariant.

\begin{prop}
\label{prop:invgbialg}
Let $\eta_{i j}$ be an invertible matrix such that $\eta_{i j}=0$ if
$|i|\neq |j|$. Consider the graded commutative bialgebra $\qg{M}(m|n)$
and impose the relations (\ref{eq:invrel1}) and
(\ref{eq:invrel2}). They are compatible with the coalgebra structure
so that we obtain again a graded commutative bialgebra
$\qg{Inv_\eta}(m|n)$. Furthermore, $\qg{Inv_\eta}(m|n)$ is a graded
Hopf algebra with antipode given by (\ref{eq:invantip}).
\end{prop}

\begin{ex}
Let $r,s$ be natural numbers. Define an invertible matrix
$\eta_{i j}$ of rank $2r+s$ by
\begin{gather*}
 \eta_{2i-1,2i}=1,\quad \eta_{2i,2i-1}=-1,\quad \text{for }
 i\in \{1,\cdots, r\},\\
 \eta_{j,j}=1,\quad \text{for }
 j\in \{m+1,\dots,m+n\},
\end{gather*}
and all other entries zero. The graded commutative Hopf algebra
$\qg{Inv_\eta}(2r|s)$ is called the \emph{ortho-symplectic supergroup}
$\qg{OSp}(2r|s)$. Its even commutative quotient is
\[
 \qg{OSp}(2r|s)\tos \qg{Sp}(2r)\tens\qg{O}(s) .
\]
\end{ex}

\begin{ex}
$\qg{OSp}(2|1)$ is the matrix super-bialgebra $\qg{M}(2|1)$ with
additional relations
\begin{gather*}
u_{1 3} u_{2 3} - u_{3 1} u_{3 2}=0, \\
2 u_{3 2} u_{3 1} + u_{3 3} u_{3 3}=1,\\
u_{1 1} u_{2 2} - u_{1 2} u_{2 1} + u_{1 3} u_{2 3}=1, \\
- u_{1 1} u_{3 2} + u_{1 2} u_{3 1} + u_{1 3} u_{3 3}=0,\\
- u_{2 1} u_{3 2} + u_{2 2} u_{3 1} + u_{2 3} u_{3 3}=0, \\
u_{2 2} u_{1 3} - u_{1 2} u_{2 3} - u_{3 2} u_{3 3}=0,\\
- u_{2 1} u_{1 3} + u_{1 1} u_{2 3} + u_{3 1} u_{3 3}=0 .
\end{gather*}
It has an antipode given by
\[
\antip \begin{pmatrix}
u_{1 1} & u_{1 2} & u_{1 3} \\
u_{2 1} & u_{2 2} & u_{2 3} \\
u_{3 1} & u_{3 2} & u_{3 3}
\end{pmatrix}
= \begin{pmatrix}
u_{2 2} & - u_{1 2} & - u_{3 2} \\
- u_{2 1} & u_{1 1} & u_{3 1} \\
u_{2 3} & - u_{1 3} & u_{3 3}
\end{pmatrix} .
\]
\end{ex}

Let us now consider the corresponding quantum groups (i.e.,
cotriangular Hopf algebras). According to Theorem~\ref{thm:cqtrbos} and
Proposition~\ref{prop:ctrbos} the cotriangular bialgebra with the same
representation category as $\qg{M}(m|n)$ is the bosonization
$\qg{M}(m|n)\rbos\CZ2$. It is generated by $\qg{M}(m|n)$ and $\CZ2$ as
an algebra with cross relations $g u_{i j}=(-1)^{|i|+|j|} u_{i j} g$
and has coproduct
\[
 \cop u_{i j}= \sum_k u_{i k} g^{|k|+|j|}\tens u_{k j},\quad
  \cop g=g\tens g .
\]
If we want to consider $\qg{M}(m|n)\rbos\CZ2$ as a symmetry of a
quantum field theory with Bose-Fermi statistics we need to perform the
spin-statistics reduction (Section~\ref{sec:rec},
Definition~\ref{def:ssred}) in order to eliminate representations with 
the wrong spin-statistics relation. In fact, this is already suggested
by our construction. The fact that the vector space $V$ has $m$ odd
and $n$ even basis vectors and not the other way round is
``forgotten'' by the super matrix-bialgebra $\qg{M}(m|n)$ as
$\qg{M}(m|n)\cong \qg{M}(n|m)$. However, writing the coaction of the
bosonization $\qg{M}(m|n)\rbos\CZ2$ on $V$ we find
\begin{equation}
 v_i\mapsto \sum_j u_{i j} g^{|j|}\tens v_j .
\label{eq:coactv}
\end{equation}
This suggest that we should in the commutative quotient
(\ref{eq:commquot}) interpret $\qg{M}(m)$ as generating the
spin. Then (\ref{eq:coactv}) precisely reflects the fact that $V$ has
the right spin-statistics relation. Conversely, the spin-statistics
reduction of $\qg{M}(m|n)\rbos\CZ2$ is given by its sub-bialgebra
generated by $u_{i j} g^{|j|}$.

Apart from the physical motivation, we can also motivate this
reduction purely mathematically by demanding that we want to consider
the universal object coacting on $V$.

\begin{prop}
The spin-statistics reduction $\qg{M}'(m|n)$ of $\qg{M}(m|n)\rbos\CZ2$
is given by its sub-bialgebra generated by the elements
$t_{i j}\defeq u_{i j}g^{|j|}$. Explicitly, it is generated by the
matrix coalgebra $\{t_{i j}\}$ with relations given by the
cotriangular structure
\[
 \cR(t_{i j}\tens t_{k l})= (-1)^{|i|\cdot |k|}
  \delta_{i j} \delta_{k l} .
\]
Explicitly,
\[
 t_{i j} t_{k l}= (-1)^{|i|\cdot |k|+|j|\cdot |l|}
 t_{k l} t_{i j} .
\]
It has a commutative quotient
\[
 \qg{M}'(m|n)\tos \qg{M}'(m)\tens \qg{M}(n)
\]
by $t_{i j}\mapsto 0$ for $|i|\neq |j|$. This is a map of cotriangular
bialgebras where $\qg{M}'(m)$ has cotriangular structure
(\ref{eq:ctrcov}) and $\qg{M}(n)$ has trivial cotriangular structure.
Furthermore, it is a (bialgebra) superextension in the sense of
Definition~\ref{def:sext}.
\end{prop}

Note that choosing the ``wrong'' spin-statistics relation also yields
a sub-bialgebra of $\qg{M}(m|n)\rbos\CZ2$ which is generated by
$\{u_{i j}g^{|j|+1}\}$. It is isomorphic to $\qg{M}'(n|m)$.

Let us now consider the spin-statistics reduction of the bosonization
of a graded Hopf algebra $\qg{Inv}_\eta(m|n)\rbos\CZ2$ that leaves an
isomorphism $\eta:V^*\to V$ invariant. In fact, instead of
constructing first
$\qg{Inv}_\eta(m|n)$ and then its spin-statistics reduction we can
proceed directly. Thus, the dual space $V^*$ carries naturally a
coaction $v^*_i\mapsto \sum_j \antip t_{j i}\tens v^*_j$ in terms of
the antipode. Now, an isomorphism $\eta:V^*\to V$ as above implies
for the antipode
\begin{equation}
\antip t_{i j}=\sum_{k,l}
  \eta_{j k} t_{k l} \eta^{-1}_{l i} .
 \label{eq:cinvantip}
\end{equation}
and thus the relations
\begin{gather}
\eta_{k l} \one = \sum_{i,j}
 \eta_{i j} t_{j l} t_{i k} ,\label{eq:cinvrel1}\\
\eta^{-1}_{k l} \one = \sum_{i,j}
 t_{l j} t_{k i} \eta^{-1}_{i j} .
 \label{eq:cinvrel2}
\end{gather}
Alternatively, these are obtained from
(\ref{eq:invantip}--\ref{eq:invrel2}) by using $u_{i j}=t_{i j}
g^{|j|}$ and the commutation relation with $g$ in
$\qg{M}(m|n)\rbos\CZ2$.

\begin{prop}
\label{prop:invcbialg}
Let $\eta_{i j}$ be an invertible matrix such that $\eta_{i j}=0$ if
$|i|\neq |j|$. Consider the cotriangular bialgebra $\qg{M}'(m|n)$
and impose the relations (\ref{eq:cinvrel1}) and
(\ref{eq:cinvrel2}). They are compatible with the coalgebra structure
and cotriangular structure so that we obtain again a cotriangular
bialgebra $\qg{Inv}_\eta'(m|n)$. Furthermore, $\qg{Inv}_\eta'(m|n)$ is
a cotriangular Hopf algebra with antipode given by (\ref{eq:invantip}).
\end{prop}

We denote the cotriangular Hopf algebra version of the
ortho-symplectic supergroup by $\qg{OSp}'(2r|s)$. Its commutative
quotient gives rise to the superextension
\begin{equation}
 \qg{OSp}'(2r|s)\tos \qg{Sp}'(2r)\tens\qg{O}(s) .
\label{eq:ospquot}
\end{equation}
Thus, physically, the spin is attached to the symplectic group
$\qg{Sp}'(2r)$. Note however, that we can construct a second version
$\qg{OSp}''(2r|s)$ based on $\qg{M}'(s|2r)$ which has a commutative
quotient
\[
 \qg{OSp}''(2r|s)\tos \qg{Sp}(2r)\tens\qg{O}'(s) .
\]
In this case the spin is attached to the orthogonal group
$\qg{O}'(s)$.

\begin{ex}
\label{ex:osp21}
$\qg{OSp}'(2|1)$ is the cotriangular matrix bialgebra $\qg{M}'(2|1)$
with additional relations
\begin{gather*}
t_{1 3} t_{2 3} + t_{3 1} t_{3 2}=0,\\
2 t_{3 1} t_{3 2} + t_{3 3} t_{3 3}=1,\\
t_{1 1} t_{2 2} - t_{1 2} t_{2 1} + t_{1 3} t_{2 3}=1,\\
t_{1 1} t_{3 2} - t_{1 2} t_{3 1} + t_{1 3} t_{3 3}=0,\\
t_{2 1} t_{3 2} - t_{2 2} t_{3 1} + t_{2 3} t_{3 3}=0,\\
t_{2 2} t_{1 3} - t_{1 2} t_{2 3} + t_{3 2} t_{3 3}=0,\\
t_{2 1} t_{1 3} - t_{1 1} t_{2 3} + t_{3 1} t_{3 3}=0 .
\end{gather*}
Its antipode is given by
\[
\antip \begin{pmatrix}
t_{1 1} & t_{1 2} & t_{1 3} \\
t_{2 1} & t_{2 2} & t_{2 3} \\
t_{3 1} & t_{3 2} & t_{3 3}
\end{pmatrix}
= \begin{pmatrix}
t_{2 2} & - t_{1 2} & t_{3 2} \\
- t_{2 1} & t_{1 1} & - t_{3 1} \\
t_{2 3} & - t_{1 3} & t_{3 3}
\end{pmatrix} .
\]
\end{ex}

\begin{ex}
\label{ex:ospsphere}
Consider the quantum homogeneous space given by
$\pi:\qg{OSp}'(2|1)\tos \qg{U}'(1)$. This is another version
of the super-sphere (see \cite{LaMa:extsuper}) which we call the
$OSp(2|1)$ super-sphere.
Here, $\pi$ is defined as the
composition of the extension map (\ref{eq:ospquot}) with
$\id\tens\cou$ and subsequently with (\ref{eq:surj}).
A set of generators of this subalgebra $\qg{S}^{2|1}$ of
$\qg{OSp}'(2|1)$ is given by
$\{t_{i 3}, t_{i 1} t_{j 2}\}$. (The relations are as in
Example~\ref{ex:osp21}.)

Analogous to Example~\ref{ex:semisphere} we can view
$(\qg{OSp}'(2|1),\qg{S}^{2|1},\qg{U}'(1))$ as the spin-bundle over
$\qg{S}^{2|1}$.
\end{ex}

For example super-Anti-de-Sitter space can be constructed precisely in
this way. As Anti-de-Sitter space is the homogeneous space
$SO(3,2)/SO(3,1)$ we pass to the spin groups
$Spin(3,2)/Spin(3,1)$ where $Spin(3,2)=Sp(4)$. Then we consider the
``physical'' quantum groups $\qg{Sp}'(4)\tos \qg{Spin}'(3,1)$ and the
superextension $\qg{OSp}'(4|1)\tos \qg{Sp}'(4)$. The corresponding
superextension of Anti-de-Sitter space is thus the quantum homogeneous
space $\qg{OSp}'(4|1)^{\qg{Sp}'(4)}$.

\subsection{The Super-Poincar\'e Group}
\label{sec:spoinc}

In this section we consider the standard super-Poincar\'e group which
is yet another type of superextension. Its
presentation here in the quantum geometric framework has some
novel aspects (in particular, the consideration of the ``physical''
quantum group version).

Recall the context of Example~\ref{ex:poinc}.
We start with the proper quantum mechanical version of the Lorentz
group $\qg{Spin}(3,1)$ which is equipped with the 
cotriangular structure
$\cR(t_{i j}\tens t_{k l})=-\delta_{i j}\delta_{k l}$ (same if one or
both $t$ carry a bar) of Bose-Fermi
statistics (see Section~\ref{sec:recpoinc}) and which we denote by
$\qg{Spin}'(3,1)$. Correspondingly, the quantum mechanical
Poincar\'e group is the quantum group $\poincs=\mink\rbos
\qg{Spin}'(3,1)$. Dually, we consider $\ealg'(\llor)$ and
$\ealg'(\lpoinc)$ (see Examples~\ref{ex:lpoinc} and \ref{ex:lsu2p}).

We are now ready to construct superextensions. We start by
considering the 4-dimensional comodule of $\qg{Spin}'(3,1)$ with basis
$\{\theta_+,\theta_-,\overline{\theta_+},\overline{\theta_-}\}$ and
left coaction in the obvious way. It gives rise to a braided
commutative comodule Hopf
algebra $\Theta_4$ as in Proposition~\ref{prop:semiext}. Dually we can
consider a 4-dimensional module of $\ealg'(\llor)$ with basis
$\{Q_+,Q_-,\overline{Q_+},\overline{Q_-}\}$ and action given as in
Example~\ref{ex:extsu2}. We can view it as
the envelope $\ealg(\lalg{\omega_4})$ of the ``abelian'' super-Lie
algebra $\lalg{\omega_4}$ with the $Q$'s forming its basis (i.e.\ $Q_i
Q_j=-Q_j Q_i$).
Both give immediately rise to semidirect superextensions
$\Theta_4\rbos\qg{Spin}'(3,1)$ and
$\ealg(\lalg{\omega_4})\rbos\ealg'(\llor)$. Furthermore, by the
induced
coaction of $\poincs$ respectively the induced action of
$\ealg'(\lpoinc)$ we obtain semidirect superextensions
$\Theta_4\rbos\poincs$
and $\ealg(\lalg{\omega_4})\rbos\ealg'(\lpoinc)$.

However, the usual Poincar\'e super-Lie algebra and supergroup are
obtained as follows.

\begin{ex}
Consider the graded commutative
$\qg{Spin}'(3,1)$-comodule Hopf
algebra $\smink$ built on the tensor product $\mink\tens\Theta_4$ and
defined as follows. It
has the tensor product algebra structure and the coalgebra structure
as for $\Theta_4$ and
\[
\cop x^\mu = x^\mu\tens 1 + 1\tens x^\mu
+ \sum_{i,j}\overline{\theta_i}\sigma^\mu_{i j}\tens\theta_j +
\sum_{i,j}\theta_i\overline{\sigma^\mu_{i j}}\tens \overline{\theta_j} .
\]
on $x^\mu$. The antipode is given by
$\antip\theta_i=-\theta_i$ and $\antip x^\mu=-x^\mu$.

$\smink$
is precisely (the algebra of functions on) the super-translation group.
The super-Poincar\'e group is now obtained analogous to the ordinary one,
namely as the bosonization $\spoincs=\smink\rbos\qg{Spin}'(3,1)$.
Explicitly, it
is generated as an algebra by $\{t_{i j},\overline{t_{i
j}}, x^\mu, \theta_\pm,\overline{\theta_\pm}\}$. It has the relations
$\theta_i \theta_j=-\theta_j \theta_i$ and $\theta_i t_{j k}=-t_{j k}
\theta_i$ and all other relations commutative. It has the matrix
coalgebra structure on $T$ while for $x^\mu$ and $\theta_i$ it is
given by
\begin{gather*}
\cop\theta_i = \theta_i\tens 1 + \sum_j t_{i j}\tens\theta_j, \\
\cop x^\mu = x^\mu\tens 1 + \sum_\nu\Lambda^{\mu \nu}\tens x^\nu
+ \sum_{i,j,k}\overline{\theta_i}\sigma^\mu_{i j}
 t_{j k}\tens\theta_k + \sum_{i,j,k}
 \theta_i\overline{\sigma^\mu_{i j}}\overline{t_{j k}}\tens
 \overline{\theta_k} .
\end{gather*}

Note that $\smink$ is at the same time a quantum homogeneous space (see
Remark~\ref{rem:bosqhom}) and carries a left coaction of $\spoincs$ as
an algebra by construction. It is thus nothing but super-Minkowski
space and gives rise to a quantum principal bundle
$(\spoincs,\smink,\qg{Spin}'(3,1))$ which we can view as its
spin-bundle (see the remarks in Examples~\ref{ex:semisphere} and
\ref{ex:ospsphere}).
\end{ex}

In fact, $\smink$ can be
viewed as a braided cocycle extension
\begin{equation}
\Theta_4\toi \smink\tos \mink
\label{eq:minkseq}
\end{equation}
(in the category of $\qg{Spin}'(3,1)$-comodules)
with the injection and surjection defined in the obvious ways. The
cocycle $\mink\to\Theta_4\tens\Theta_4$ which determines the extension
is given by
\begin{equation}
x^\mu\mapsto
\sum_{i,j}\overline{\theta_i}\sigma^\mu_{i j}\tens\theta_j +
\sum_{i,j}\theta_i\overline{\sigma^\mu_{i j}}\tens \overline{\theta_j} .
\label{eq:minkcocy}
\end{equation}
For details of
the relevant Hopf algebra extension theory we refer the
reader to \cite{Maj:qgroups}. (The generalization to the $Z_2$-grading
is straightforward in the present case.)

We can equally utilize the enveloping algebra picture.
The sequence (\ref{eq:minkseq}) becomes after dualization and
restriction to the (super) Lie algebras
\[
 \mom\toi \smom\tos\omega_4 .
\]
Now $\smom$ is a central (graded) extension of $\omega_4$ by
$\mom$. Here, the action of $\ealg'(\llor)$ on $\omega_4$ is given as
in Example~\ref{ex:extsu2}, with the same action for the barred
generators and the action between barred and un-barred generators zero.
The extension is determined by the graded cocycle
$\omega_4\tens\omega_4\to \mom$ given by
\[
 \overline{Q_i}\tens Q_j\mapsto 2 \sum_\mu \sigma^\mu_{i j} P^\mu .
\]
This is the familiar way to look at the
super-translation Lie algebra $\smom$, see e.g.\ \cite{AzIz:lie}.
The enveloping Hopf algebra version of this is
precisely the dual of (\ref{eq:minkcocy}) and lives in the category of
$\ealg'(\llor)$-modules.

\begin{ex}
$\smom$ is the $\ealg'(\llor)$-module super-Lie algebra built on the
space $\omega_4\oplus\mom$ with bracket
\[
 \{\overline{Q_i},Q_j\}=2 \sum_\mu \sigma^\mu_{i j} P^\mu
\]
and all brackets involving $P^\mu$ vanishing.
The ordinary super-Poincar\'e Lie algebra is now the semidirect
product $\smom\rtimes\llor$. However, the
proper quantum group (determining the physical symmetries) is the
``enveloping'' Hopf algebra
$\ealg'(\lspoinc)\defeq\ealg(\smom)\rbos\ealg'(\llor)$. Note
that it differs from $\ealg(\smom\rtimes\llor)$ which can be
constructed as a graded enveloping Hopf algebra in that it takes into
account the spin-statistics relation as explained above.
The cross-relations between $\{E,F,H\}$ and $\{Q_\pm\}$ are as in
Example~\ref{ex:extsu2} and extended to the barred generators in the
obvious way. The coproducts for all generators are primitive, except
for $\cop Q_i = Q_i\tens 1 + \xi\tens Q_i$.

Due to Lemma~\ref{lem:pbos} $\spoincs$ and $\ealg'(\lspoinc)$ are
dual (co)triangular Hopf algebras. This induces a natural quantum
tangent space
which can be identified with $\lspoinc$ (i.e., it is the smallest
quantum tangent space that contains $\lpoinc$). The derivatives are
the ones of Example~\ref{ex:derivpoinc} supplemented by
\begin{gather*}
 \partial_{H}(\theta_\pm)=\pm\theta_\pm, \quad
 \partial_{\overline{H}}(\overline{\theta_\pm})
 =\pm\overline{\theta_\pm}, \quad
 \partial_{E}(\theta_+)=\theta_-, \quad
 \partial_{\overline{E}}(\overline{\theta_+})
 =\overline{\theta_-}, \\
 \partial_{F}(\theta_-)=\theta_+, \quad
 \partial_{\overline{F}}(\overline{\theta_-})
 =\overline{\theta_+}, \quad
 \partial_{Q_i}(\theta_j)=\delta_{i j},\quad
 \partial_{\overline{Q_i}}(\overline{\theta_j})=\delta_{i j},\\
 \partial_{Q_i}(x^\mu)=\sum_j \overline{\sigma^\mu_{i j}}
  \overline{\theta_j},\quad
 \partial_{\overline{Q_i}}(x^\mu)=\sum_j \sigma^\mu_{i j}
  \theta_j .
\end{gather*}
All other derivatives of generators vanish.
\end{ex}

\section{Conclusions and Outlook}

We have exhibited here a categorial point of view on quantum
field theory yielding a generalized notion of symmetry based on quantum
groups and braided categories.
This is motivated by the observation that rather than symmetry groups
themselves, only their representation categories
are operationally relevant in quantum field theory.
The resulting framework unifies the concepts of conventional
symmetry and exchange statistics (as was already noticed in
\cite{Oe:spinstat}). We have shown how (super)group
symmetry, Bose-Fermi statistics and the spin-statistics relation 
are interconnected in a three-layer structure that recovers
the generalized quantum group symmetry of quantum field theory.

Rephrasing the old question of non-trivially extending space-time
symmetries in the new framework naturally leads to supersymmetry
(assuming Bose-Fermi statistics).
Furthermore, we were able to show that (in this framework)
supersymmetry
is indeed the most general way of unifying external and internal
symmetries. Even if we drop the non-triviality condition only group
symmetries and supersymmetries are allowed.
This appears to be a no-go theorem for
``hidden'' (non-triangular) quantum group symmetries in physically
interesting theories such as the standard model.

However, this has to be interpreted with care.
A crucial ingredient in our formulation is the condition that the
quantum group which extends the given space-time-statistics quantum
group does not modify the statistics. (That is, the cotriangular
structure is preserved by the extension).
We see this as a natural constituent of the extension problem (and it
is implicit in its conventional formulation).
For example, it would be conceivable that there exist
multiplets of states with braid statistics of which so far only
(bosonic or fermionic) singlets have been observed. But as this goes
in a sense beyond ordinary quantum field theory it also goes beyond
our formulation of the extension problem.
Furthermore, the braiding is defined for all objects in the
relevant category while not all of them can be necessarily interpreted
as being subject to some exchange statistics in the conventional
sense. It is thus conceivable that the braiding can have a broader
meaning in general and just reduce on the relevant objects to the
conventional statistics.
This would also leave open the possibility for non-triangular quantum
group symmetries.

We have also seen that for theories
with non-symmetric braid statistics (e.g.\ anyons in two spatial
dimensions), the separation
between the 
conventional notions of symmetry and statistics can no longer be
retained (see the end of
Section~\ref{sec:qgsym} and the beginning of Section~\ref{sec:frec}).
Only the generalized notion of quantum group
symmetry remains applicable.
It is thus no surprise that (non-triangular) quantum groups are indeed
employed in the construction of
fractional supersymmetry (which implies non-symmetric braid
statistics) \cite{DMAP:geofsusy,Dun:braidfsusy}.

Finally, we mention that there is a generalization of
quantum field theory \cite{Oe:bqft}
to precisely the categorial framework (braided
categories) we outline in Section~\ref{sec:qgsym}. This
naturally takes in quantum group symmetries and for the proper
Poincar\'e quantum group reconstructed in Section~\ref{sec:recpoinc}
yields automatically
the correct differences for path integrals and Feynman
rules between bosons and fermions \cite{Oe:spinstat}.
In this context the present paper clarifies how supersymmetric
theories would have to be constructed in this framework (namely through
their proper quantum group versions considered here).
Furthermore, even $q$-deformations of supersymmetries (of which some
examples have been considered in the literature, see e.g.\
\cite{Man:qdefsuper}) can thus be employed. 
As $q$-deformation has been proven to be a potential
regulator of quantum field theory \cite{Oe:bqft} this yields the prospect
of a (notoriously difficult) covariant regularization of
supersymmetric theories. 

\section*{Acknowledgements}

I would like to thank H.~Pfeiffer, T.~Sch\"ucker and F.~Girelli for their
careful reading and many helpful comments on the manuscript.
This work was supported through a NATO postdoctoral fellowship.

\bibliographystyle{amsordx}
\bibliography{stdrefs}
\end{document}